\documentclass[10pt,fleqn,a4paper]{article}
\pdfoutput=1
\usepackage{graphicx}
\usepackage{epsfig,cite}
\usepackage{amssymb}
\usepackage{amsmath}
\usepackage{color}
\usepackage{amstext,alltt,setspace}
\usepackage{amsbsy}
\usepackage{array}
\usepackage{hyperref}
\usepackage{slashed}
\usepackage{url}
\usepackage{geometry}
\begin{document}

\emergencystretch=.5em

\begin{flushright}
Edinburgh 2021/19\\
  TIF-UNIMI-2021-5\\
\end{flushright}

\vspace*{.2cm}

\begin{center}
  {\Large \bf{Correlation and Combination of Sets of Parton Distributions}}
\end{center}

\vspace*{.7cm}

\begin{center}
Richard D.~Ball$^{1}$,    Stefano Forte$^{2}$ and Roy Stegeman$^{2}$
  \vspace*{.2cm}

{  \noindent
  {\it
$^1$The Higgs Centre for Theoretical Physics,\\
 University of Edinburgh, JCMB, KB, Mayfield Rd, Edinburgh EH9 3FD, Scotland\\
        ~$^2$ Tif Lab, Dipartimento di Fisica, Universit\`a di Milano and\\
        INFN, Sezione di Milano,
        Via Celoria 16, I-20133 Milano, Italy\\
}}

      \vspace*{3cm}

      {\bf Abstract}
\end{center}

{\noindent
We study the correlation between different sets of parton distributions (PDFs).
Specifically, viewing different PDF sets as distinct determinations, 
generally correlated,
of the same underlying physical quantity, we examine the extent to which the
correlation between them 
is due to the  underlying data. We do this both for
pairs of PDF sets determined using a given fixed methodology, and between sets
determined using different methodologies. We show that correlations have a
sizable component that is not due to the underlying data, because the data 
do not determine the PDFs uniquely.
We show that the data-driven correlations
can be used to assess the
efficiency of methodologies used for PDF determination. We also show that the
use of data-driven correlations for the combination of different PDFs into a
joint set can  lead to inconsistent results, and thus that the
statistical combination used in constructing the widely used PDF4LHC15
PDF set remains the most reliable method.}

\pagebreak

\section{Uncertainties and correlations of parton distributions}
\label{sec:intro}
Correlations between parton distribution functions (PDFs) are an important
ingredient in the accurate determination of PDFs and their
uncertainties, notoriously a current major challenge of
contemporary collider precision phenomenology~\cite{Gao:2017yyd}.
The correlation
between PDFs and physical observables is  a tool used
to understand both the impact of PDF uncertainties on
predictions, and the impact of individual observables on PDF
determination, as originally emphasized in
Ref.~\cite{Nadolsky:2008zw}, and subsequently exploited for Higgs phenomenology
in Sect~3.2 of Ref.~\cite{Dittmaier:2012vm}. However, more
fundamentally, correlations are an integral part of the determination
of PDF uncertainties.

Indeed, just as full information on data
uncertainties requires knowledge of the experimental covariance matrix, and not
just the uncertainty on individual datapoints, full information on PDF
uncertainties is encoded in the point-by-point covariance matrix
between all pairs of PDFs at any pair of points:
\begin{equation}\label{eq:covmatpdfgen}
 {\rm Cov}[f_a^p,f_a^q](x,x')
  =E[ f_a^p(x,Q_0^2)f_a^q(x',Q_0^2)]
  -E[ f_a^p(x,Q_0^2)] E[ f_a^q(x',Q_0^2)] ,
\end{equation}
and associated correlation matrix
\begin{equation}\label{eq:corrpdf}
  \rho[f_a^p,f_a^q](x,x')
  =\frac{ {\rm Cov}[f_a^p,f_a^q](x,x')}{\sqrt{{\rm Var}[f_a^{p}]{(x)}{\rm Var}[f_a^{q}]{(x')}}},
\end{equation}
where $f^p_a(x,Q_0^2)$ is the $p$-th PDF with momentum fraction $x$ and a chosen reference scale
$Q^2_0$ in a given PDF set $\Phi_a$ and we denote by ${\rm
  Var}[f^{p}_a]$ the variance, i.e. the diagonal element of the
covariance matrix
\begin{equation}\label{eq:varpdf}
  {\rm Var}[f^{p}_a](x)= {\rm Cov}[f_a^p,f_a^p](x,x)
  =E[ f_a^p(x,Q_0^2)^2] -E[ f_a^p(x,Q_0^2)]^2.
\end{equation}
In what follows, we will use indices $p,q$ to label PDF flavors, and
indices $a,b$ to label PDF sets. So $p$ runs over gluon, up, anti-up, down,
\dots, while $a$ runs over NNPDF3.1, NNPDF4.0, MSHT20, CT18\dots.
Henceforth, we will
generally consider the case in which $x=x'$ and
suppress the $x$ and $Q_0^2$ dependence, so for instance we will write
the covariance Eq.~(\ref{eq:covmatpdfgen}) as 
\begin{equation}\label{eq:covmatpdfgen1}
{\rm Cov}[f_a^p,f_a^q]\equiv  {\rm Cov}[f_a^p,f_a^q](x,x) 
  =E[ f_a^p f_a^q]
  -E[ f^p_a] E[ f_a^q] .
\end{equation}
In Eqs.~(\ref{eq:covmatpdfgen}-\ref{eq:covmatpdfgen1}), $E$ denotes the average
over the probability distribution
of PDFs; so  the PDF
covariance matrix $ {\rm Cov}[f_a^p,f_a^q](x,x')$ is the second central moment of the
joint probability distributions of PDFs.
It can be computed in a standard way~\cite{Butterworth:2015oua}
given a representation of this probability distribution, specifically
as a multigaussian in parameter space for a given PDF parametrization,
or as a Monte Carlo sample of PDF replicas.

The correlation between PDF flavors Eq.~(\ref{eq:corrpdf})
(F-correlation, henceforth)  is a standard
concept, and it has been widely computed and used. However, one may
also define a different kind of correlation: the correlation
between PDF sets~\cite{HERAFitterdevelopersTeam:2014fzy,froidevauxtalk:note}. Each PDF set is  a determination of
the same  underlying true PDFs. As such, each set can be viewed as a
different determination of the same underlying physical quantity. Generally, two
determinations of the same quantity are characterized both
by an uncertainty, and a correlation between them. In the presence of
uncertainties, each determination may be thought of as a random
variable. So for a distinct pair of determinations one can define 
their covariance and correlation, which may then be combined using the
standard
methodology that is used for the combination of correlated measurements~\cite{dagos,Cowan:1998ji}.
Indeed, the
correlation between two determinations expresses the amount of new
information that each determination introduces:
two completely
independent determinations are fully uncorrelated, while two completely
correlated determinations are simply repetitions of the same determination.

Hence we can view any particular PDF set  $\Phi_a$ containing PDFs 
$f^p_a(x,Q_0^2)$ and their uncertainties as an
instance of a probability distribution of PDF
determinations~\cite{Giele:2001mr}, just 
like any measurement is an instance of a probability distribution of
measurement outcomes. The covariance matrix Eq.~(\ref{eq:covmatpdfgen1}) can
then be viewed as a special case of a more general covariance matrix
\begin{equation}\label{eq:selfcovmatpdf}
 {\rm Cov}[f_a^p,f_b^q]\equiv {\rm Cov}[f_a^p,f_b^q](x,x) =E[ f_{a}^p f^q_{b}] -E[  f_{a}^p] E[ f_{b}^q] 
\end{equation}
and corresponding correlation 
\begin{equation}\label{eq:selfcorrpdf}
  \rho[f_a^p,f_b^q]
  =\frac{{\rm Cov}[f_a^p,f_b^q]}{\sqrt{{\rm Var}[f_a^{p}]{\rm Var}[f_b^{q}]}},
\end{equation}
where again, for simplicity, we only consider the case $x=x'$ and
suppress the dependence of the PDFs on $x$ and $Q_0^2$.

Equations~(\ref{eq:selfcovmatpdf},\ref{eq:selfcorrpdf})
provide the covariance and correlation, generally between two
different PDF flavors, across two
different PDF sets, $\Phi_a$ and $\Phi_b$. We will therefore refer to
them as cross-covariance and cross-correlation, respectively.
The average in Eq.~(\ref{eq:selfcovmatpdf})
is now  performed over the full probability
distribution of PDF sets, i.e. of distinct determinations, of which sets
$\Phi_{a}$ and  $\Phi_{b}$ are two generic instances. How this might
be done in practice is something that we will discuss below. 

It is clear that if we take $p=q$ the cross-covariance and cross-correlation
Eqs.~(\ref{eq:selfcovmatpdf},\ref{eq:selfcorrpdf}) reduce to the
covariance and correlation between
the two different determinations of PDF $f^p$ provided by sets
$\Phi_a$ and $\Phi_b$. Indeed, in this case
\begin{equation}
  \label{eq:selfcovmatpdfdiag}
 {\rm Cov}[f_a^p,f_b^p]=E[ f_{a}^pf^p_{b}] -E[
  f_{a}^p] E[ f_{b}^p] ,
\end{equation}
and the corresponding correlation is
\begin{equation}\label{eq:selfcorrpdfdiag}
  \rho[f_a^p,f_b^p]
  =\frac{{\rm Cov}[f_a^p,f_b^p]}{\sqrt{{\rm Var}[f_a^p]{\rm Var}[f_b^p]}}.
\end{equation}
Equations~(\ref{eq:selfcovmatpdfdiag},\ref{eq:selfcorrpdfdiag}) provide
respectively the covariance and correlation between two different determinations
$f^p_a(x,Q_0^2)$, $f^p_b(x,Q_0^2)$ of the $p$-th PDF at point $x$, viewed as two
different determinations of the underlying true value, i.e. the
covariance and correlation between PDF sets, or S-covariance and S-correlation.

The F-correlation and S-correlation are both special cases of the
cross-correlation: the former, specializing to the case in which one
compares two different PDFs from the same PDF set, and the latter, specializing to the
case in which one compares the same PDF  in two
different PDF sets. The F-covariance is a covariance in a space of
PDFs, within a given set. The S-covariance is instead the covariance
in a space of different PDF sets, which is a rather less trivial
concept. Specifically, if a Monte Carlo representation is adopted, then
a PDF set  $\Phi_{a}$ is represented as a set of Monte Carlo PDF
replicas  $\{f_{a}^{p,(r)}(x,Q_0^2)\}$, and 
the F-covariance Eq.~(\ref{eq:covmatpdfgen}) can be computed by
simply averaging over these replicas. In order to achieve something similar
for the S-covariance, one would first need to construct
replicas that span the space of
possible independent determinations of a given PDF --- including, say,
the results that might have been found by different 
groups using different methodologies. This is clearly nontrivial.

 Elucidating the nontrivial aspects  of averaging over the  space of
PDF determinations  is 
the main goal of this paper. 

The reason why the structure of the space of PDF determinations is subtle 
is that
the outcome of a PDF determination does not only depend on the
underlying data. Indeed, a PDF determination amounts to the
determination of  a
probability distribution in a space of functions~\cite{Giele:2001mr}, 
using as input  a discrete set of datapoints: hence the outcome  is
not unique. Specifically, if a very general functional form, such as a
neural network,  is adopted, then there is an ensemble of best fits of
equal quality to fixed underlying data. It has indeed been shown
explicitly~\cite{Ball:2014uwa}  in the NNPDF framework that if  the same data
are fitted over and over again, a distribution of best-fit PDFs  is
obtained,  rather than a single answer. In an approach such as the
CT18 PDF determination~\cite{Hou:2019efy} a
more restrictive parametrization is adopted and a
unique best fit is found, but then the  fit is
repeated with a large number underlying functional forms also leading
to a distribution of best fits.

It is clear that the distribution of best fits for fixed underlying
data provides a contribution to the uncertainty (and more generally to
the covariance matrix) that is unrelated to the data uncertainty.
Accordingly,  when
computing the S-correlation  Eq.~(\ref{eq:selfcorrpdfdiag}) there
is a contribution to it which comes from integrating over the space of
PDFs which correspond to fixed underlying data. But of course, the
correlation by construction lies between $-1$ and $+1$, so this also
implies that inevitably the data-induced component of the S-correlation is
strictly less then one, because it does not include the
component that is not data-driven.
We will refer to the non-data-driven component as ``functional
correlation'' for short (without committing ourselves to its precise
origin). Note that a priori the non-data-driven component also
contains uncertainties related to theory choices, such as the
value of $\alpha_s$, or missing higher order corrections. These are
however in principle (though not necessarily in practice) easy to 
account for by simply varying the relevant parameters, and we will
not discuss them here: we will always consider theory assumptions to
be fixed. 

In the remainder of this paper, we will investigate the relative sizes
of the data-driven and functional components. This will be done by
computing  explicitly the data-driven S-correlation for PDFs determined
from the same underlying data.
The question we will address is then: how large is the functional component of
the correlation, in comparison to the data-driven one --- negligible, sizable,
perhaps even dominant?

This question is not purely academic, but in fact quite important for
applications. First, as mentioned, the F-correlation
Eq.~(\ref{eq:corrpdf}) is routinely used to estimate the impact of
a given dataset on the PDFs. But then, surely knowledge of the size of
the data-driven component of the S-correlation is necessary for such an
assessment, because it tells us how much the PDF is determined by the
underlying data.
Furthermore, correlations between independent measurements are
normally used when combining different measurements of the same
quantity~\cite{dagos}. Combined PDF sets, such as the PDF4LHC15
set~\cite{Butterworth:2015oua}, can be viewed in a similar way,
namely, as the combination of different determinations of the
true PDF. At present, these combinations are performed by simply
assuming that all PDF sets in the combination are equally
likely. However, one might think that they should instead be combined
as correlated measurements, and that a determination of the
data-driven S-correlation might be useful to this
goal~\cite{froidevauxtalk:note}.
It is then interesting to investigate possible ways to implement such a 
procedure, and their consequences.

We will answer these questions by computing the data-induced
S-correlation explicitly. In Sect.~\ref{sec:self}
we will present results for the data-driven component of the
S-correlation, both  between different PDF
sets determined from the same underlying data and with the same
methodology, and between pairs of PDF sets determined
using the same data, but different methodologies. We will use the
results to shed light on the origin of the
PDF S-correlation, and we will explain how S-correlations can be used as
a diagnostic tool when comparing different methodologies. In
Sect.~\ref{sec:comb} we will discuss the
implications of our result for the construction of combined PDF
sets. After discussing the PDF4LHC15 prescription for the combination of PDF 
sets in Sect.~\ref{sec:stat}, in Sect.~\ref{sec:unc} we will discuss how this
could be replaced by a correlated combination, and the consequences of
doing so.

\section{PDF cross-correlations and S-correlations}
\label{sec:self}

\subsection{Correlated replicas}
\label{sec:correp}

Our goal is to compute the cross-correlation between PDF sets, which
we can determine from the  cross-covariance
Eq.~(\ref{eq:selfcovmatpdf}). 
We assume that a Monte Carlo representation is used for PDF sets. This
assumption is not restrictive, since Hessian PDF sets can be converted
into Monte Carlo sets using the methodology of Ref.~\cite{Watt:2012tq}.
A PDF set $\Phi_a$ is then represented by a set of $N$ PDF replicas $\{{f_a^{p(r)}}: r = 1,\ldots
N\}$ of the $p$-th PDF flavor $f_a^p$, that provide an importance sampling of the probability
distribution of the the PDF set. We assume for simplicity that the
number of PDF replicas $N$ is fixed and sufficiently large that a
faithful representation of the underlying probability representation
is obtained, and we will provide an estimate of the uncertainty due to
finite-size effects. Also, we
assume that the PDF sets that are being considered all provide a
faithful representation of the underlying data, so that they would for
instance pass a closure test~\cite{Ball:2015oha,Ball:2021leu}.

Each replica is equally probable, so replicas are statistically
uncorrelated, and estimators of functions of the PDFs are given by simple
averages over the replicas,
\begin{equation}
\langle X[f_{a}^p]\rangle = \frac{1}{N}\sum_{r=1}^{N} X[{f_a^p}^{(r)}].
\label{eq:EXf}
\end{equation}
Thus in particular the mean and the F-covariance of the PDF set are given by
\begin{equation}
E[f_{a}^p] = \langle f_a\rangle, \qquad {\rm Cov}[f_{a}^p,f_{a}^q]
=\langle f_{a}^pf_{a}^q \rangle - \langle f^p_a\rangle\langle f^q_a\rangle.
\label{eq:ECovf}
\end{equation}
Henceforth, we will use angle brackets to denote the replica average,
while the symbol $E$ used in Sect.~\ref{sec:correp} denotes the generic average.

The way PDF replicas are constructed is by first constructing a Monte
Carlo representation of the underlying data, i.e. by producing an
ensemble of data replicas, such that the ensemble mean reproduces the central
value of the original data, and the  covariance over the ensemble
reproduces the data covariance matrix. These data replicas can then be
fitted with whatever methodology is chosen: i.e., a fit to each data
replica is performed thereby leading to a best-fit PDF for the given
replica. The ensemble of best fits then provides the Monte Carlo
representation of PDFs. Note that this can be done in conjunction with
any PDF fitting methodology, such as, for example, the MSTW methodology
discussed in Ref.~\cite{Watt:2012tq}.

When computing  the F-covariance
Eqs.~(\ref{eq:covmatpdfgen},\ref{eq:ECovf}), the same replica is used
to determine  $f_a^p$ and $f_a^q$, so in particular
\begin{equation}\label{eq:corrav}
\langle f_a^pf_a^q\rangle=\frac{1}{N}\sum_{r=1}^{N}{f_a^p}^{(r)}  {f_a^q}^{(r)}.
\end{equation}
It is important to understand that each PDF replica  ${f_a^p}^{(r)}$ is
obtained by fitting to an individual data replica, but, as mentioned, the result of
this fit is generally not unique. Indeed, as also mentioned, a
distribution of different PDFs is found even when fitting repeatedly
the same underlying data. In the NNPDF methodology, that uses a very
general neural network as underlying parametrization, this simply
follows from the fact that the best-fit neural network trained to
fixed underlying data is not unique: each  training to fixed
underlying data produces a
different but equally good answer. In
other methodologies this is achieved by considering an ensemble of
different functional forms fitted to the same data, each of which
gives a different best fit~\cite{Aaron:2009aa,Hou:2019efy,Bailey:2020ooq}.
Clearly, the spread
of PDFs fitted to fixed underlying data is completely unrelated to the data uncertainties and
correlations. Thus, in order to get the full variance, covariance,
${f_a^p}^{(r)}$ and
${f_a^q}^{(r)}$ in Eq.~(\ref{eq:corrav}) must be PDFs that correspond to the same
PDF replica: i.e. to one fixed  fit to  a fixed data replica.

We would now like to turn to the computation of the S-covariance
and S-correlation
between two PDF sets $\Phi_a$ and  $\Phi_b$,
that we can write  as 
\begin{align}
{\rm Cov}[f_a^p,f_b^p] &= \langle f_a^p f_b^p\rangle - \langle f_a^p\rangle\langle
f_b^p\rangle, \label{eq:xcov}\\
\rho [f_a^p,f_b^p] &=\frac{{\rm Cov}[f_a^p,f_b^p]}{\sqrt{{\rm Var}[f_{a}^p]{\rm Var}[f_{b}^p]}}. \label{eq:xcorr}
\end{align}
Naively, one could think that this might be computed by starting with two sets
of PDF replicas representing the two PDF sets, $\{{f_a^p}^{(r)}\}$ and
$\{{f_b^p}^{(r)}\}$. However it is clear that
if these PDF replicas are chosen independently,  then the
S-covariance, and thus the 
S-correlation 
Eq.~(\ref{eq:selfcorrpdfdiag})
will always vanish, because the replicas
are uncorrelated between sets. Indeed, the same applies to the
computation of the standard F-covariance Eq.~(\ref{eq:ECovf}). Namely,
if two independent sets of 
$N$ PDF replicas chosen at random from the same PDF set $\Phi_a$
are combined into a set of $2N$  independent PDF replicas
then
\begin{equation}\label{eq:corravz}\frac{1}{N}\sum_{r=1}^N {f_a^p}^{(r)}{f_b^p}^{(N+r)}=   
\langle f_a^p\rangle\ \langle f^p_b\rangle\
\end{equation}
and the S-covariance vanishes, where the equality (and thus the
vanishing) hold in the large $N$ limit, i.e. within the accuracy of
the Monte Carlo representation.

In order to compute the correlation between PDF sets we must thus
determine the  S-covariance by using PDF replicas that are
correlated between PDF set $\Phi_a$ and PDF set
$\Phi_b$. In order to understand what this means, assume for a
moment, for the sake of argument, that the data do determine uniquely
the best-fit PDF: so that for each data replica there is a unique PDF
replica. In this case, when we label each PDF replica
${f_a^p}^{(r)}$
the index $r$ corresponds uniquely to the specific data replica to
which the PDFs have been fitted.
Assuming  that the two given sets are obtained using
two different methodologies, this unique PDF best fit might be slightly
different in sets $\Phi_a$ and 
$\Phi_b$, so ${f_a^p}^{(r)}\not = {f_b^p}^{(r)}$. The 
 S-correlation between PDF sets $a$ and $b$
can be computed by simply making sure that the same correlated replica 
$\{{f_a^p}^{(r)}\}$  and $\{{f_b^p}^{(r)}\}$ is used:
the cross-covariance  
Eq.~(\ref{eq:xcov}) is then generally  non-zero, since the PDF replicas are 
correlated by the choice of the common underlying data replica $r$.
This  also automatically implies that the PDF set has
unit correlation to itself, as it must, because if $a=b$ then the
unique answer  ${f_a^p}^{(r)}$ is 
used in the computation
of averages.

However, in actual fact, we know that this is generally
not the case: the data replica does not uniquely determine
the PDF. So, in a realistic  case, for each data
replica $r$ and for each PDF set  $\Phi_a$ there is a distribution of best
fits $\{{f_a^p}^{(r,r')}\}$ where now the index $r'$ runs over
``functional'' or methodological replicas: for each data replica,
$r$, the  index $r'$ labels all other aspects that determine the
answer. For instance, if parametrization dependence is estimated by
varying the functional form, as in
Refs.~\cite{Aaron:2009aa,Hou:2019efy,Bailey:2020ooq}, then $r'$ would
label the different functional forms that are used.
If $a=b$, then for fixed $r$ and fixed $r'$ the same answer is
obtained and the correlation is one: each PDF set has unit correlation
to itself.   If $a\not =b$, then the S-correlation is found by
varying both $r$ and $r'$ in a correlated way. If however only $r$ is
correlated, and the average over $r'$ is performed in an uncorrelated
way, the corresponding component of the correlation vanishes
according to Eq.~(\ref{eq:corravz}). 

So, if only $r$ is correlated, then the computation
provides a non-vanishing component
of the S-correlation, namely, its data-driven
component, but it misses the functional component, that would also
require correlating $r'$,  and thus it does not
give the full correlation. Now, it is clear that if we take $a=b$ in
Eq.~(\ref{eq:xcov})
then the S-covariance reduces to the simple variance  and 
the S-correlation
must equal one, at least in the limit of large $N$: it is the
correlation of a PDF set to itself. However, if only the data-driven component is included in
the computation of the cross-correlation, this consistency check is not
satisfied: rather, the result may come out to 
be less than unity. This
deviation from unity tells us, crudely speaking, the extent to which the
PDFs are determined by the underlying data. The deviation from one
measures the size of the decorrelation due not to having integrated
over functional replicas in a correlated way: it tells us how
significant the functional
component of the S-correlation is.

It is interesting to ask whether the  functional component of the 
cross-correlation could
be computed directly. In order to do this, one would need ``functional
replicas'', namely, one would need to explicitly construct replicas
$\{{f_a^p}^{(r,r')}\}$, such that varying the index $r'$ fully spans
the possible best fits obtained from a fixed underlying data
replica $r$. This could perhaps  be done when varying aspects 
of the  methodology that influence the final outcome, and that are
controlled  by a parameter:
this parameter would then be fixed to take the same value
when computing the S-covariance
    Eq.~(\ref{eq:xcov}). We will see an explicit example below, when
    discussing correlated preprocessing.
However, this appears to be  nontrivial in general  
    for non-parametric methodology aspects that determine in which of
    the many equivalent best fits a particular minimization will
    end up.

    In what follows, we will focus on the computation of the
    data-driven component of the S-correlation. This means that we
    will use Eq.~(\ref{eq:xcov}), and we will view the replica index
    $r$ as an index that labels the underlying pseudodata
    replica. Then, when  comparing two different PDF sets  $\Phi_a$ and
    $\Phi_b$, the replicas  $\{{f_a^p}^{(r)}\}$  and
    $\{{f_b^p}^{(r)}\}$ are fitted to the same underlying
    data. However, we will also  ``compare a PDF set  $\Phi_a$  to itself''. By
    this, we mean that we will compute the S-correlation (or
    cross-correlation) by using in Eq.~(\ref{eq:xcov}) two sets of PDF
    replicas fitted to the same data replicas using the same
    methodology,
     $\{{f_a^p}^{(r,r')}\}$  and  $\{{f_a^p}^{(r,r'')}\}$,
    where, as above $r$ labels the data replica while the indices $r'$, $r''$
    span the space of fits to the same data replica. Hence in
    particular 
\begin{equation}\label{eq:scorrav}
\langle f_a^pf_a^p\rangle=\frac{1}{N}\sum_{r=1}^{N}{f_a^p}^{(r,r')}  {f_a^p}^{(r,r'')}.
\end{equation}
    In practice, since we have no control on the values of $r'$,
    $r''$,  $\{{f_a^p}^{(r,r')}\}$  and  $\{{f_a^p}^{(r,r'')}\}$, are
    simply two different fits from PDF set $\Phi_a$  to the same
    $r$-th underlying data replica. This is to be contrasted to
    Eq.~(\ref{eq:corrav}), used in the computation of the F-correlation, in
      which the same fit to the same data replica is used:
      effectively, Eq.~(\ref{eq:corrav}) corresponds to taking
      $r'=r''$ in Eq.~(\ref{eq:scorrav}).
    We will refer to the situation in which Eq.~(\ref{eq:scorrav}) is
    used   as comparing a PDF set to
    itself, and we will refer to the resulting correlation as the data-driven
    component of the correlation.

\begin{figure}[t]
\begin{center}
  \includegraphics[height=0.25\textwidth]{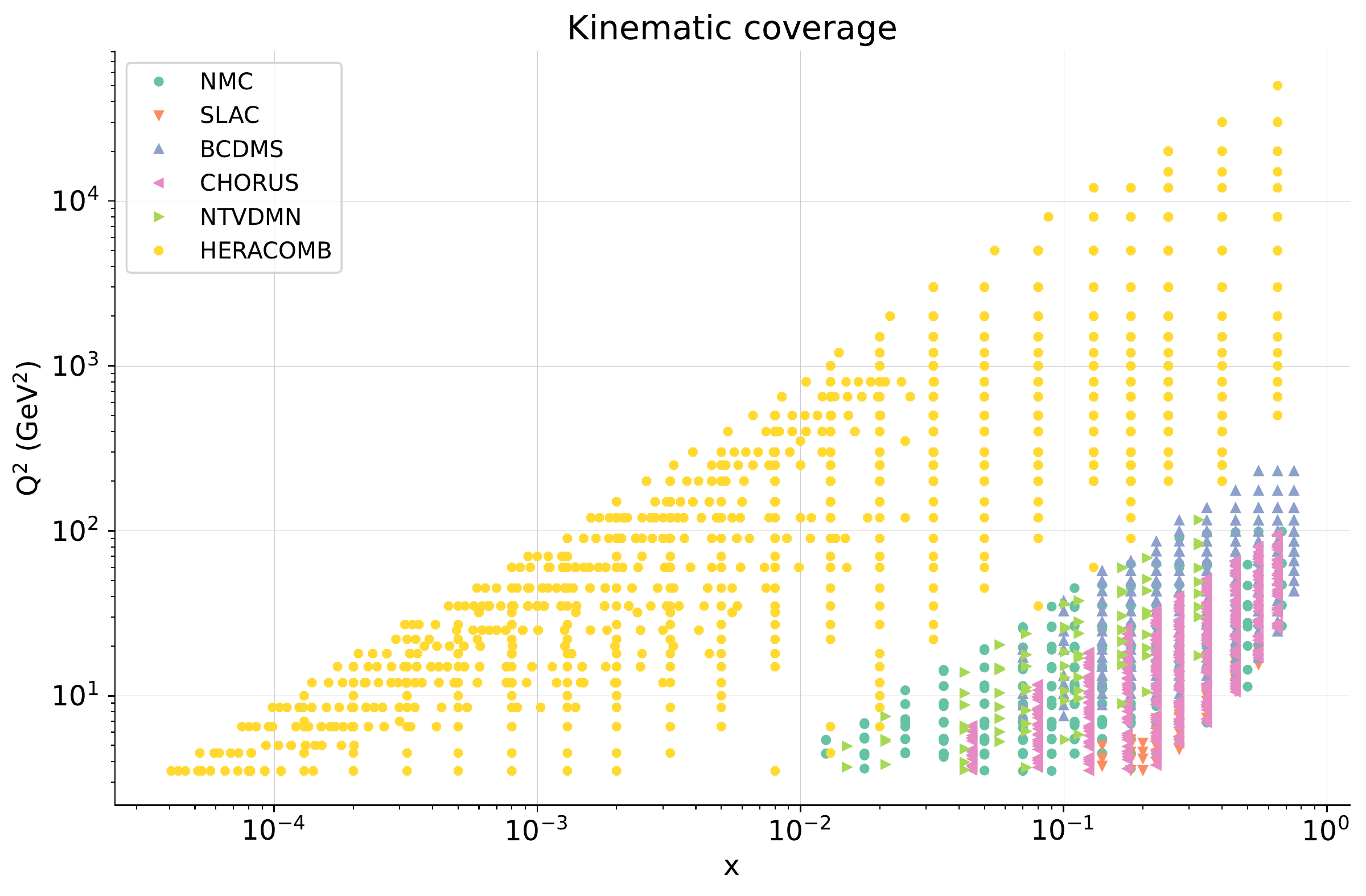}%
  \caption{\small The kinematic coverage in the $(x,Q^2)$ plane of the DIS dataset used for
    the determination of the PDF sets discussed in this paper.
    \label{fig:scatter}}
\end{center}
\end{figure}

\subsection{ The data-induced S-correlation and cross-correlation}
\label{sec:dcorr}

We compute the data-induced component of the S-correlation,
both  between a pair of PDF sets
determined from the same underlying data using two different
methodologies, and between a PDF set and itself, as defined at the end
of Sect.~\ref{sec:correp}.
Specifically, we  consider the 
NNPDF3.1 methodology as presented in
Ref.~\cite{Ball:2017nwa}, and the NNPDF4.0 methodology recently  used
for the
NNPDF4.0 PDF set~\cite{Ball:2021leu}. 
The NNPDF4.0 methodology differs from the NNPDF3.1 methodology mostly because of the choice of PDF
parametrization and minimization algorithm, and its main feature is that
it is obtained through a hyperoptimization
procedure~\cite{Forte:2020yip}. Here, the details of these two
methodologies are not relevant, and it suffices to know that they are both faithful, and compatible with
each other (as we will
verify explicitly) .

In order to compute the F-correlations and S-correlations shown in this section,
we have generated several PDF sets of approximately 1000 PDF replicas
each, using the open-source NNPDF code~\cite{NNPDF:2021uiq}.
These PDF replicas have been generated by performing fits to a dataset which
consists of  the deep-inelastic scattering (DIS)
data used for the NNPDF3.1 PDF determination, as listed in Table~1 of
Ref.~\cite{Ball:2017nwa}. The kinematic coverage of this data in the
$(x,Q^2)$ plane is shown in Fig.~\ref{fig:scatter}. This and all other
plots in this paper are produced using  \texttt{validphys}~\cite{zahari_kassabov_2019_2571601, NNPDF:2021uiq}.
The reasons for
choosing a DIS-only dataset are to limit the use of computational resources,
as well as to deal with a dataset involving a single, well-understood process, thereby avoiding
possible complications related to tensions between data, slow
perturbative convergence, and other issues that could obscure our
conclusions. Uncertainties on the S-correlations due to the
finite size of the replica sample are estimated using a bootstrapping
procedure. Further details on the computation of the S-correlation
are given in Appendix~\ref{app:correp}.

\begin{figure}[t]
\begin{center}
  \includegraphics[width=0.49\textwidth]{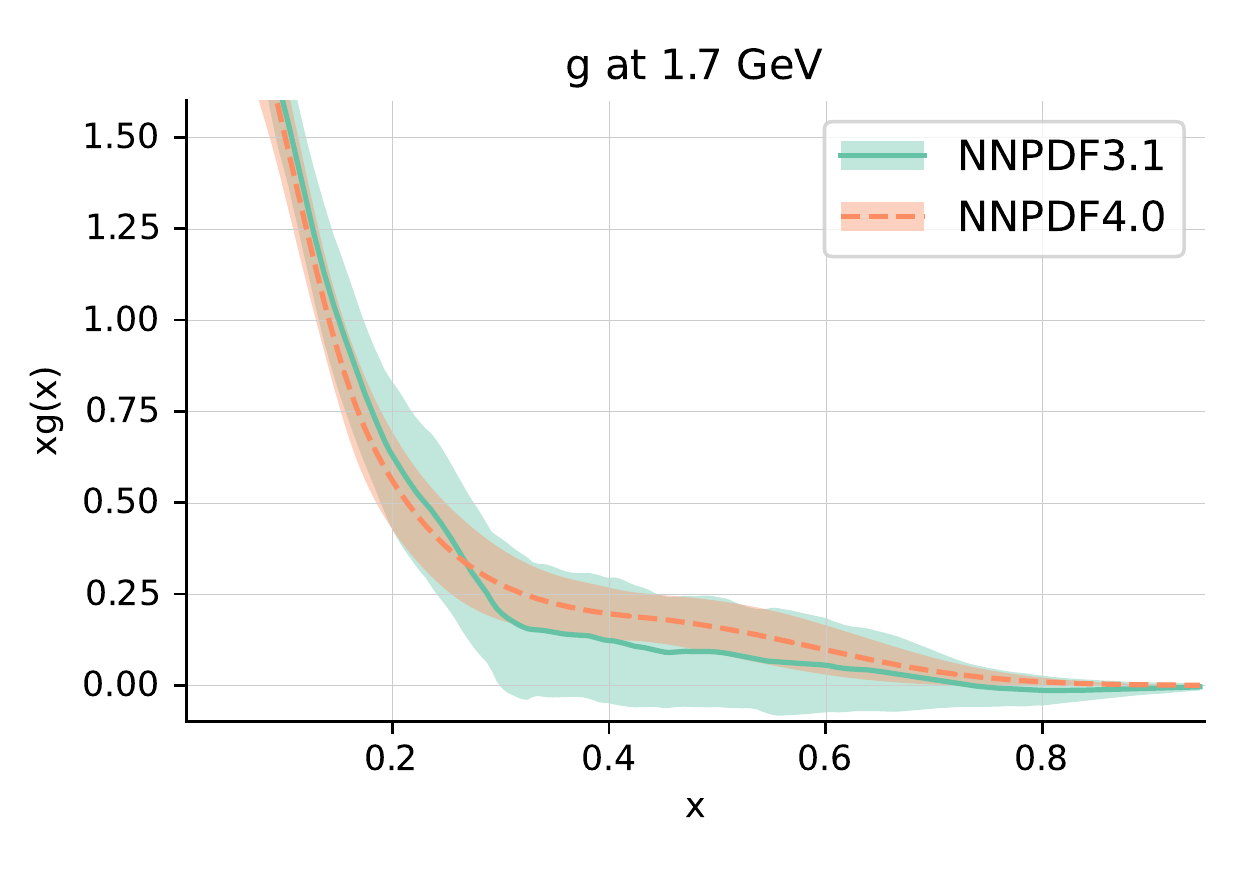}%
  \includegraphics[width=0.49\textwidth]{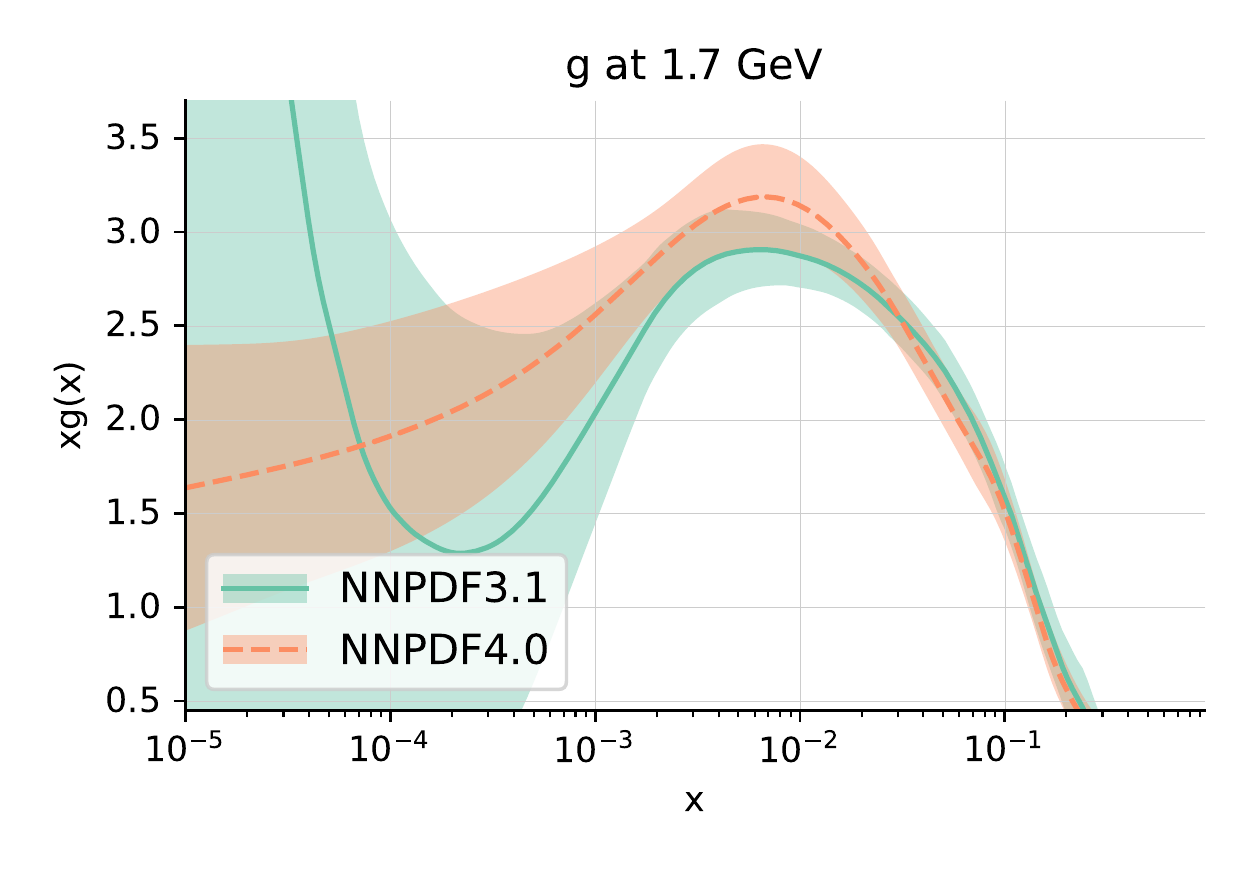}\\
  \caption{\small The gluon PDF as determined with
      the NNPDF3.1 and NNPDF4.0 methodologies, on  a linear (left)
      and logarithmic (right) scale in $x$.
    \label{fig:glucomp}}
\end{center}
\end{figure}
To begin with, we have constructed four PDF sets, all determined from the same
underlying data replicas: two
using the NNPDF3.1 methodology, and two using the NNPDF4.0 methodology. Whereas a
detailed comparison of PDFs produced using the NNPDF4.0 and NNPDF3.1
methodologies is given in Ref.~\cite{Ball:2021leu}, in Fig.~\ref{fig:glucomp} we 
show a representative comparison of the gluon PDF as determined using these
two methodologies. The general features of the comparison discussed
in Ref.~\cite{Ball:2021leu} are apparent from this example: namely,
first, that results
found with either methodology are
compatible within uncertainties and central values are generally quite
similar, and second, that the NNPDF4.0 methodology leads to rather
smaller uncertainties, so generally whereas the NNPDF4.0 central value
is within the NNPDF3.1 uncertainty band, the NNPDF3.1 central value is
not within the rather smaller NNPDF4.0 uncertainty band.

\begin{figure}[t]
\begin{center}
  \includegraphics[width=0.49\textwidth]{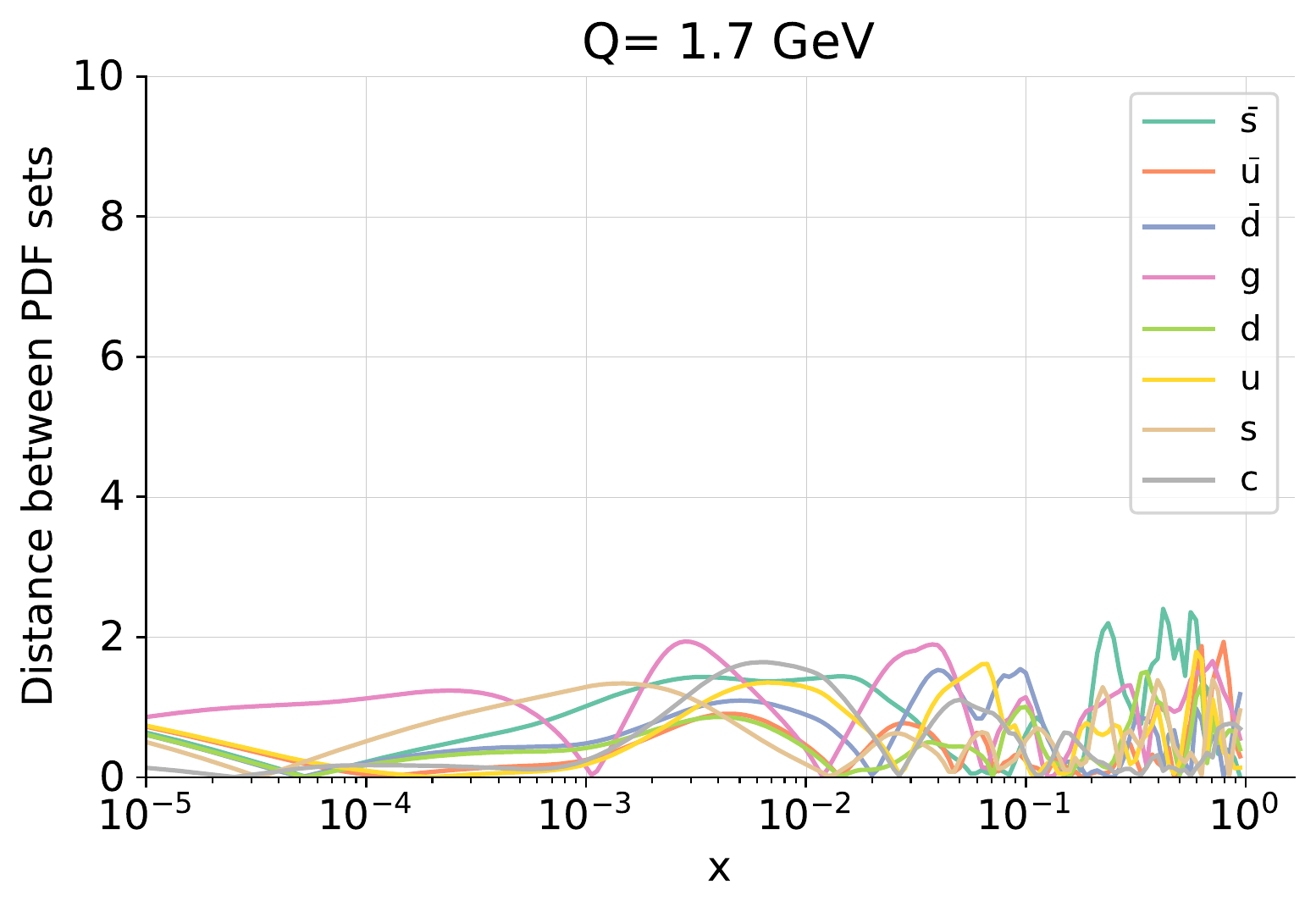}%
  \includegraphics[width=0.49\textwidth]{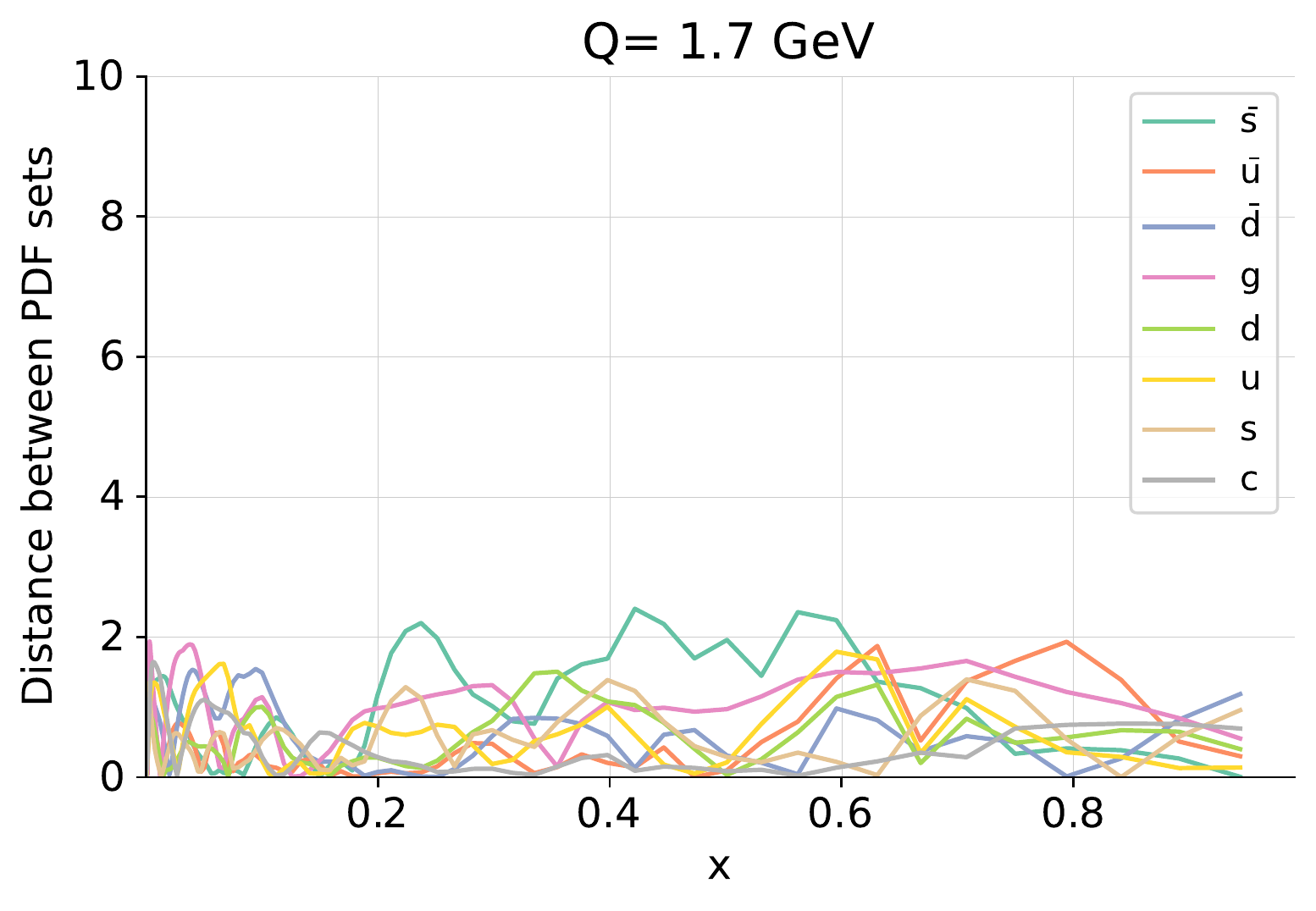}\\
  \includegraphics[width=0.49\textwidth]{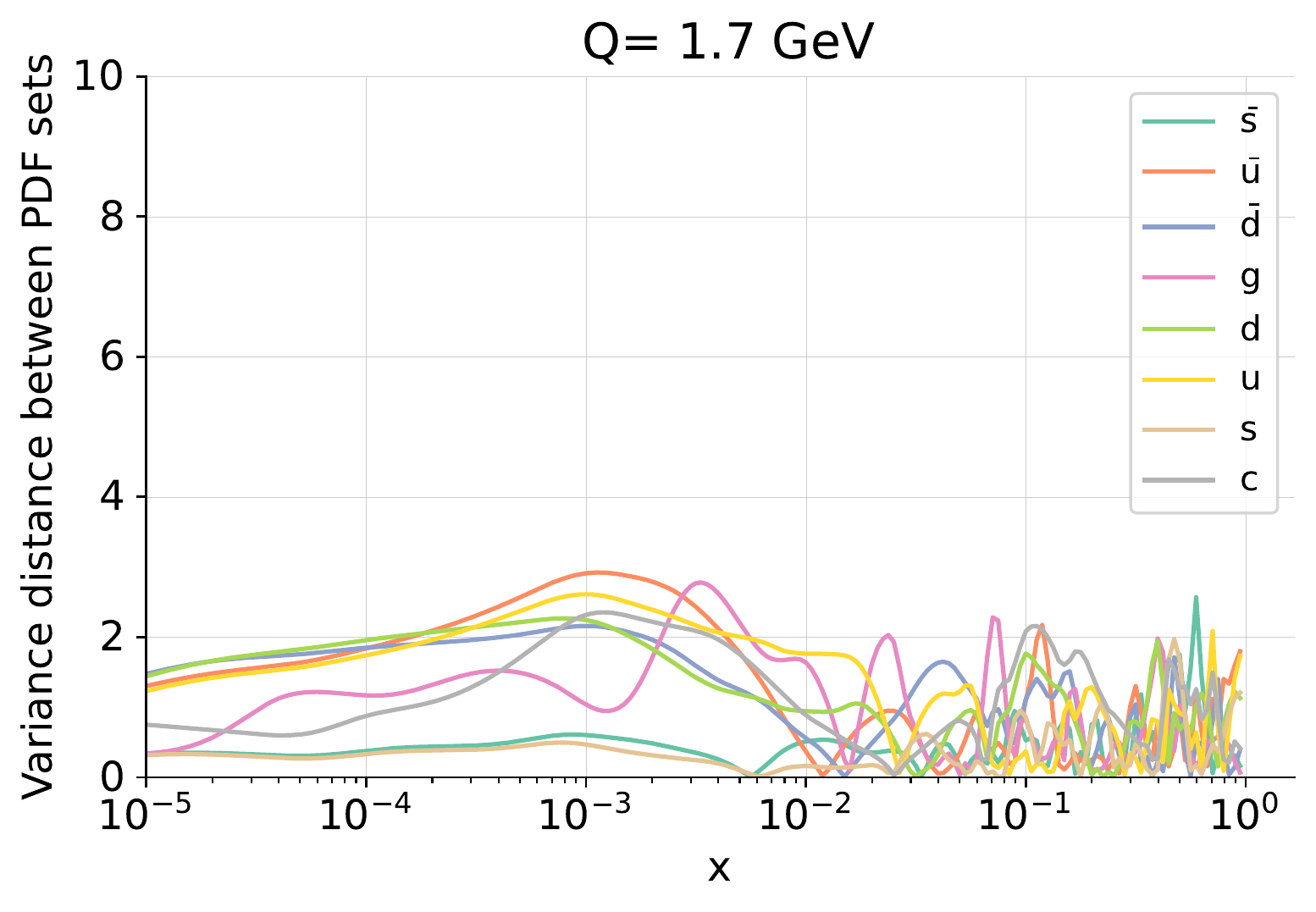}%
  \includegraphics[width=0.49\textwidth]{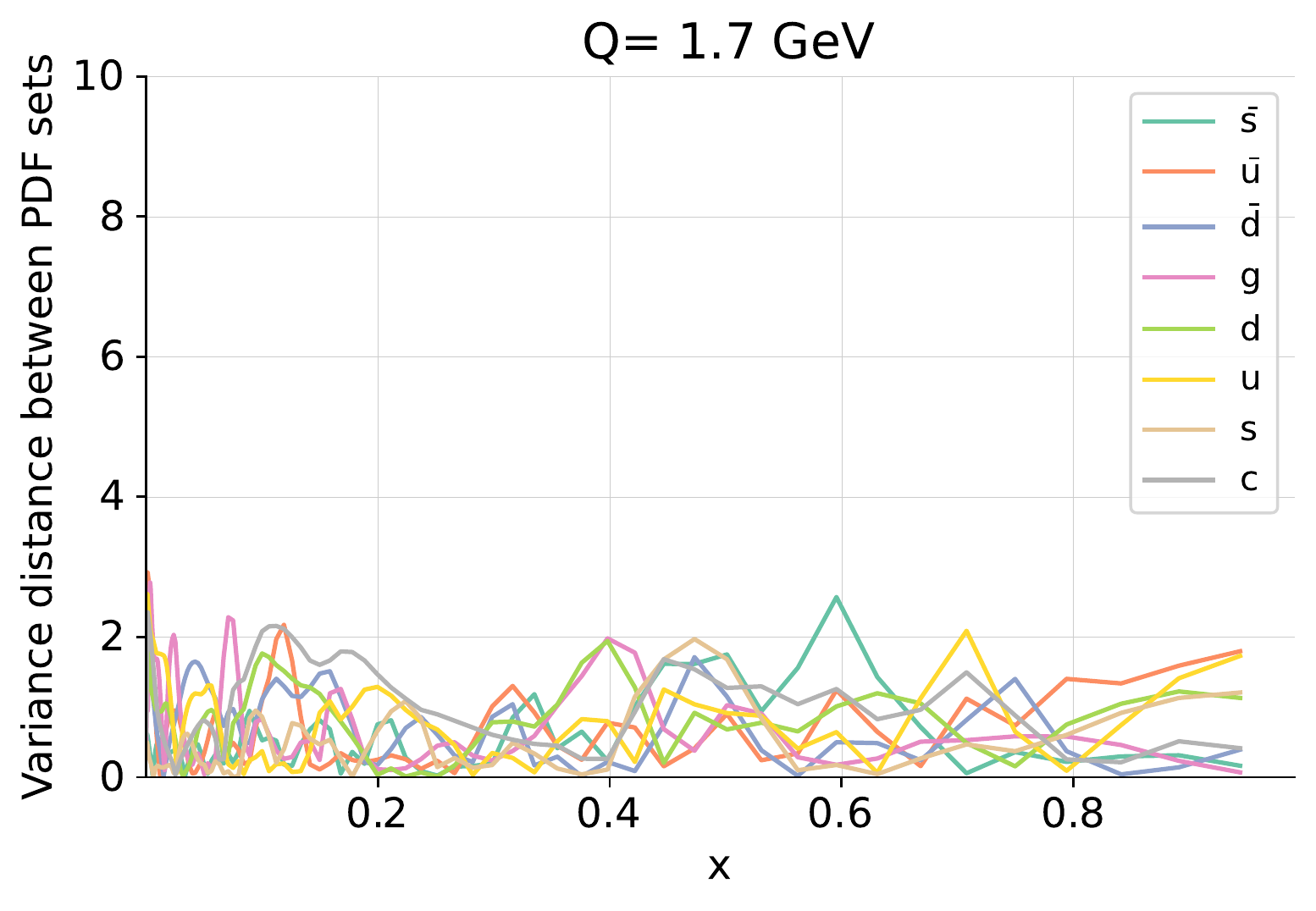}\\
  \caption{\small Distances between central values (top) and uncertainties
      (bottom) of two
      PDF sets determined using the NNPDF4.0 methodology. Results are
      shown both on a logarithmic (left) and linear (right) scale in $x$.
  \label{fig:distances}}
\end{center}
\end{figure}
As in all NNPDF determinations, it can be
checked explicitly that independently determined PDF replicas all
provide a consistent representation of the same underlying probability
distribution. Namely, we can check that the standard deviation of the mean of
$N_{\rm rep}$ replicas is equal to
$\sigma/\sqrt{N_{\rm rep}}$. In order
to perform this check, we have generated yet another set of PDF
replicas with the NNPDF4.0 methodology, now based on a new set of data
replicas. In
Fig.~\ref{fig:distances} we show the distance between the central
values and uncertainties of the two different sets of
PDF replicas determined using  the NNPDF4.0 methodology trained on different
sets of data replicas. The distance
(defined in Appendix B of Ref.~\cite{Ball:2014uwa}) is
the mean square difference of central values in units of the
standard deviation of the mean. It is apparent that indeed its value
is of order one, as it ought to be. This shows that as the number of
replicas increases, both central values and uncertainties of PDFs
converge to the mean and standard deviation of the underlying probability
distribution.

\begin{figure}[t]
  \begin{center}
    \includegraphics[width=0.49\textwidth]{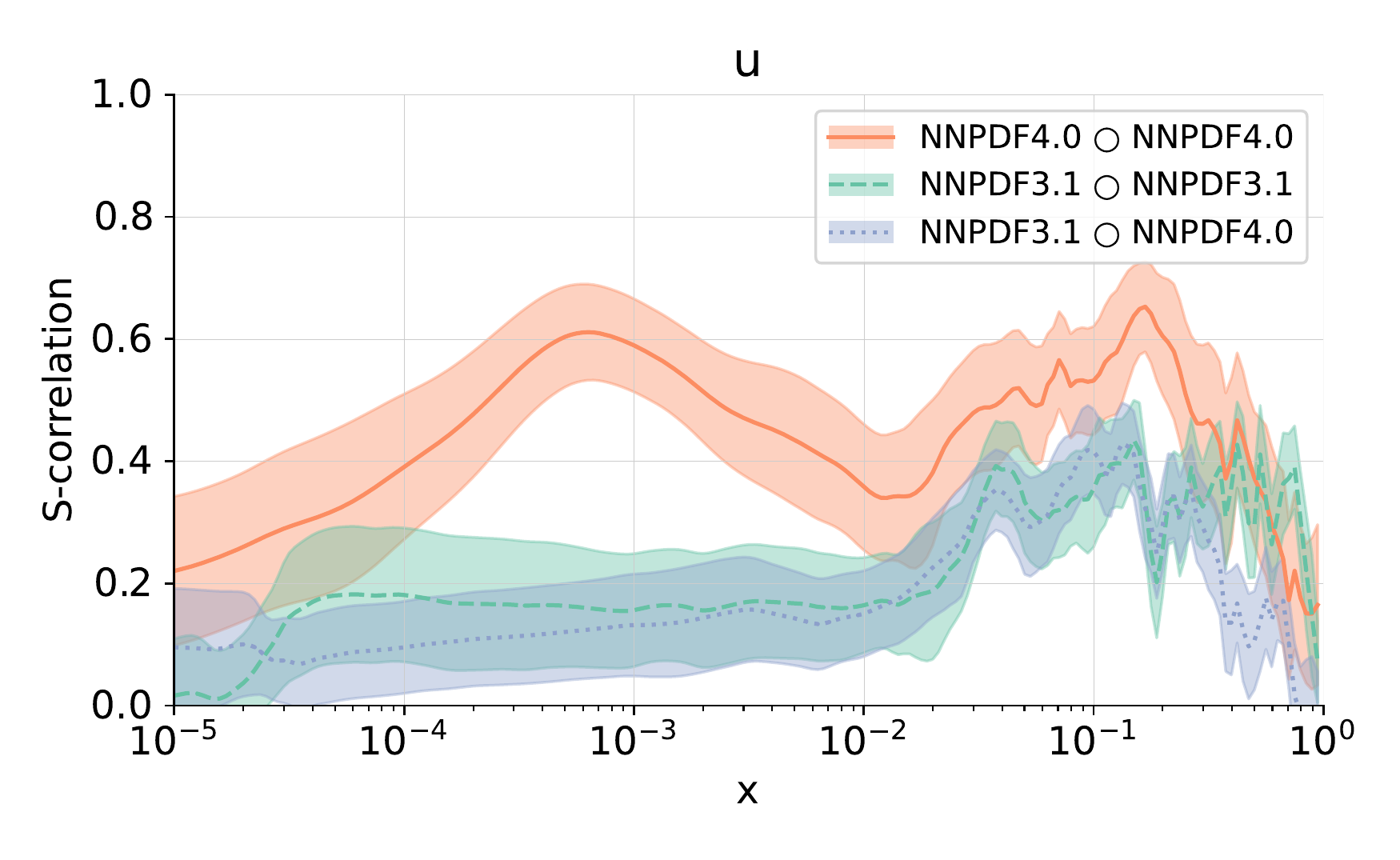}
    \includegraphics[width=0.49\textwidth]{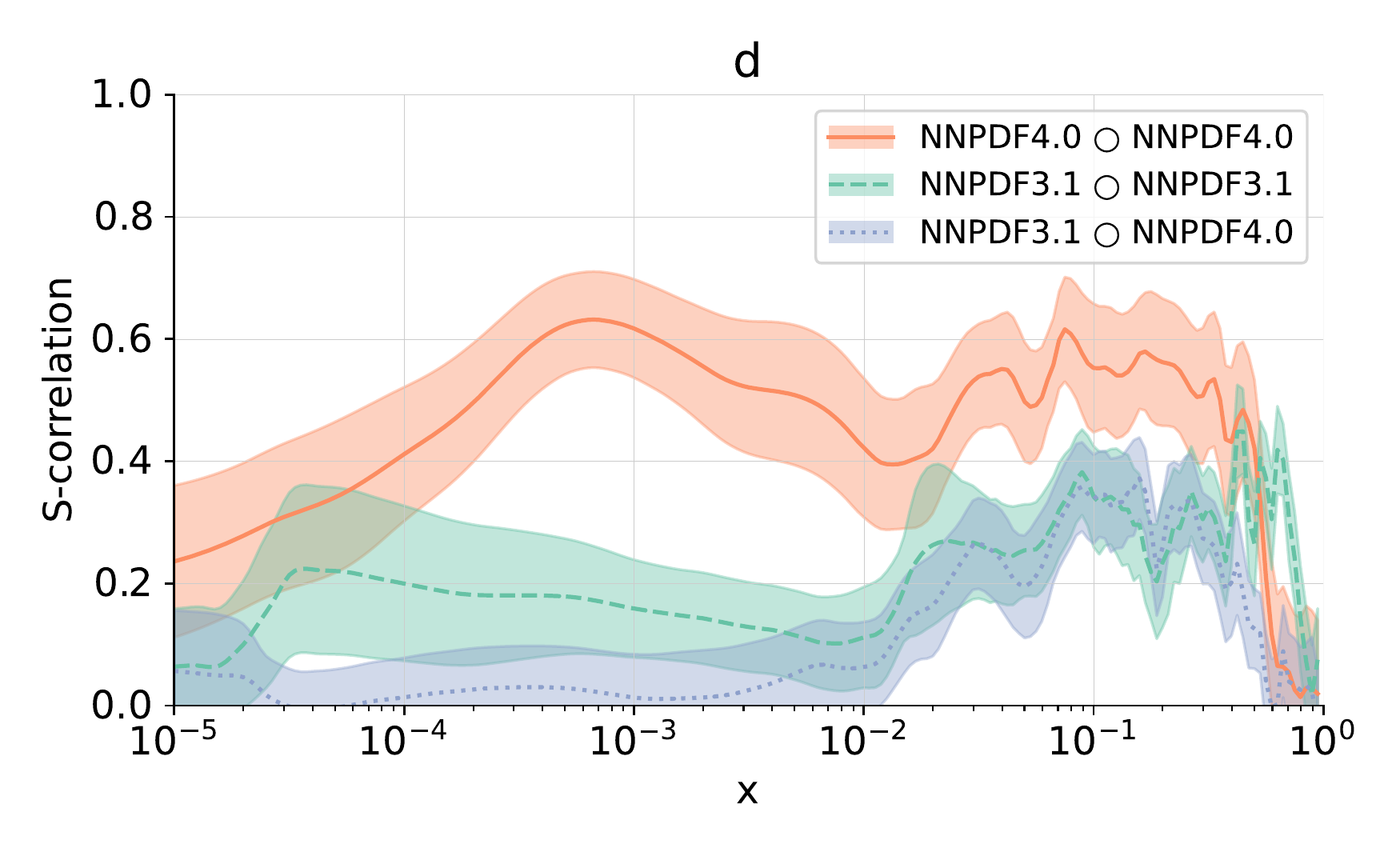}\\
    \includegraphics[width=0.49\textwidth]{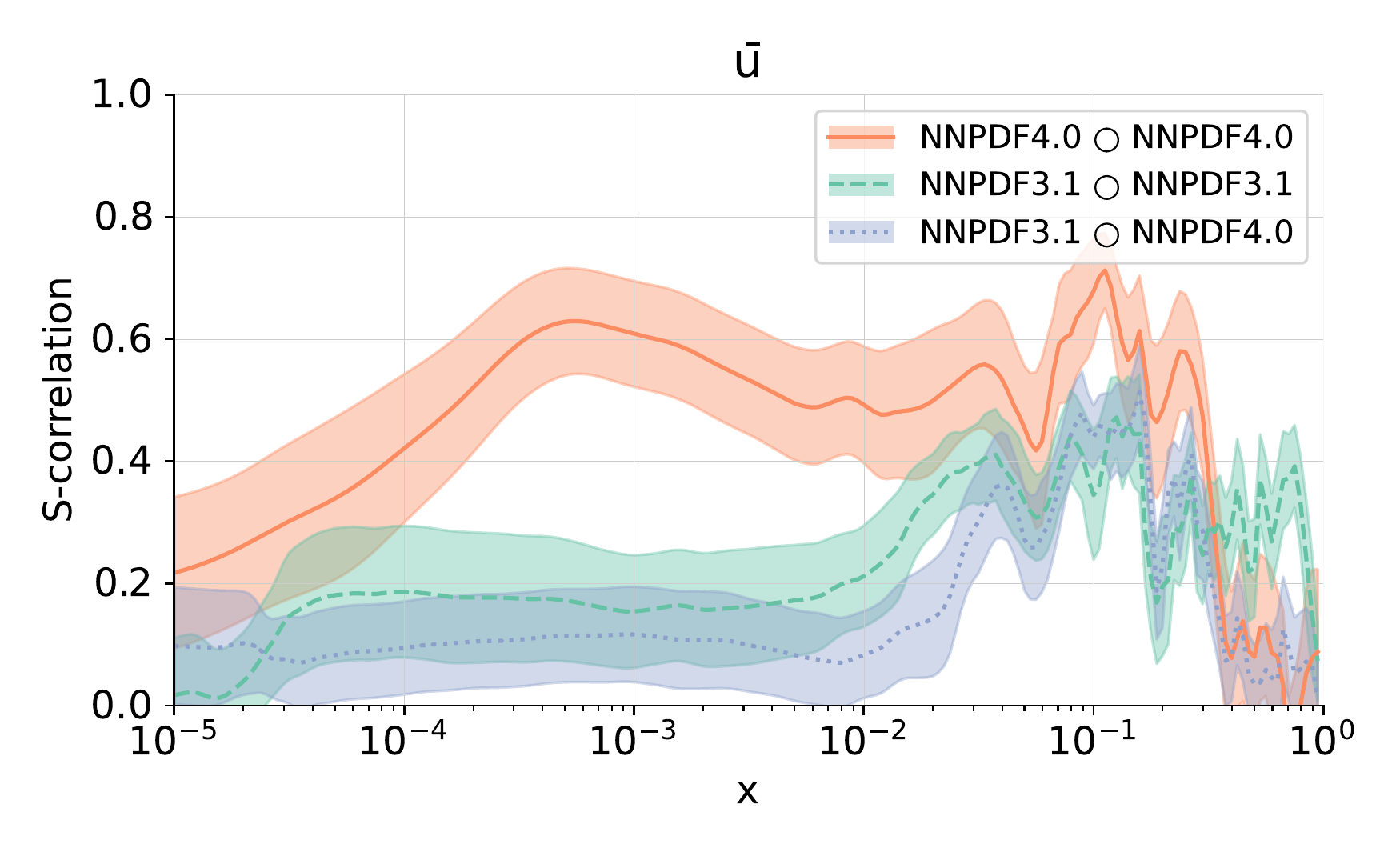}%
    \includegraphics[width=0.49\textwidth]{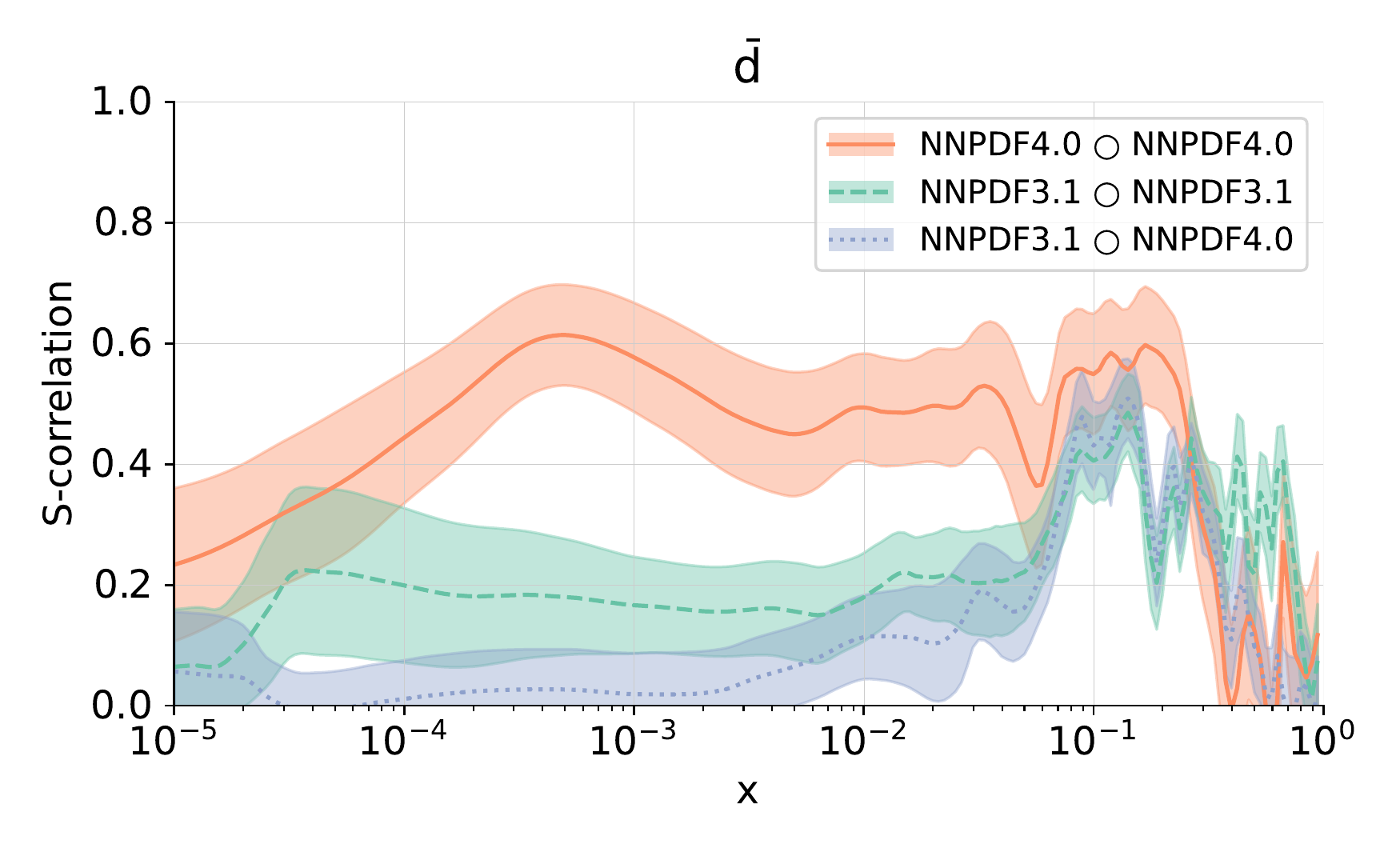}\\
    \includegraphics[width=0.49\textwidth]{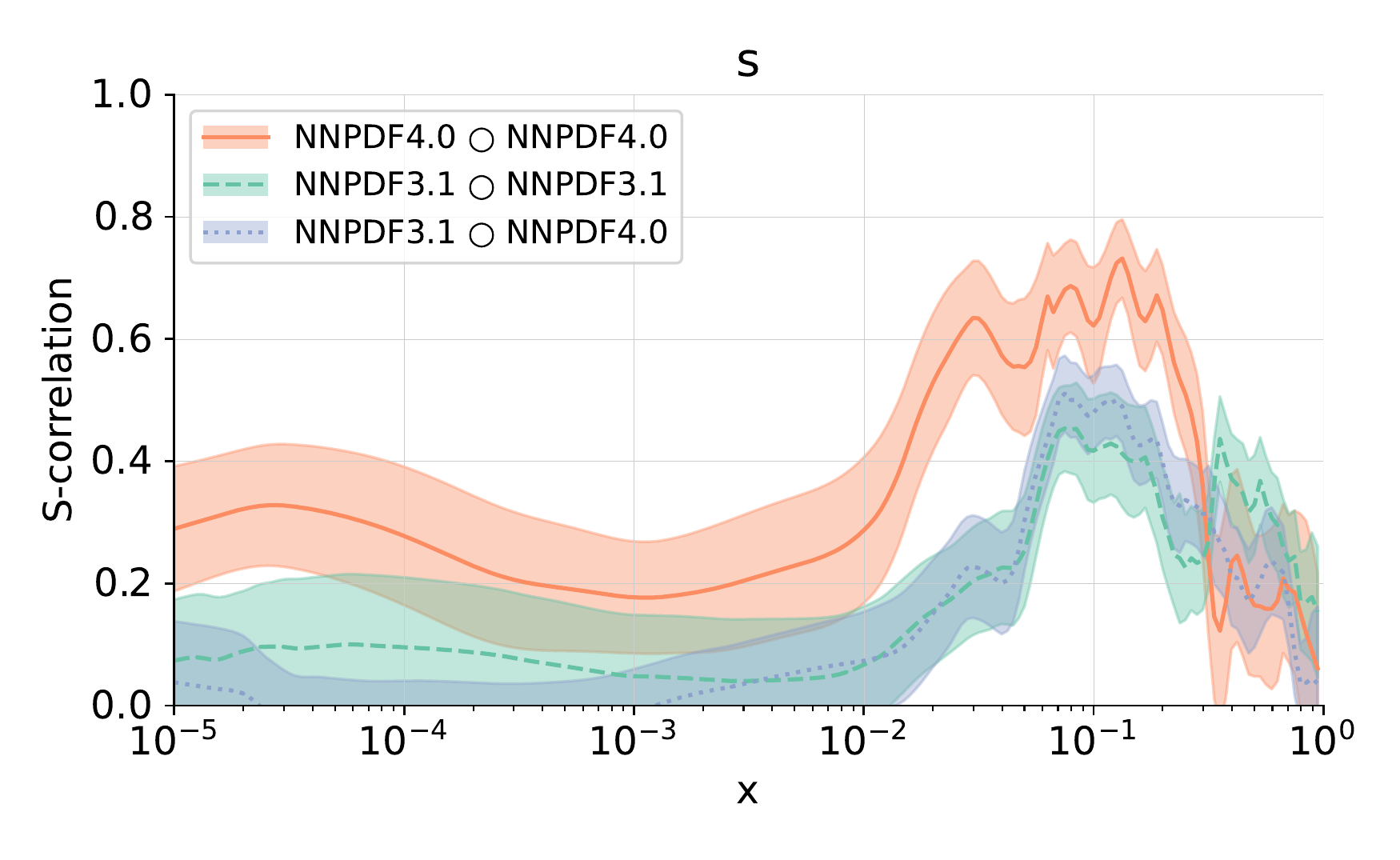}%
    \includegraphics[width=0.49\textwidth]{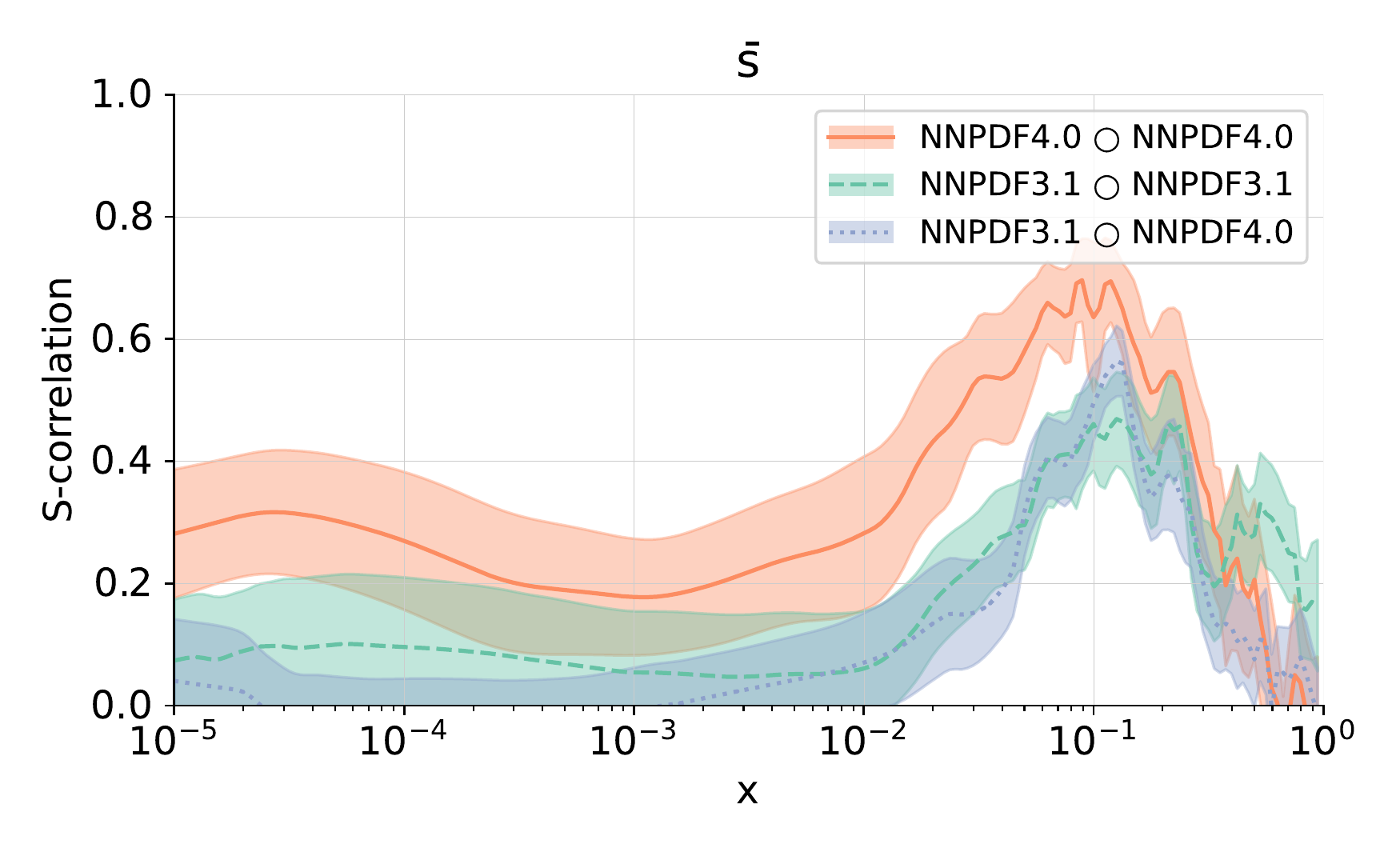}\\
    \includegraphics[width=0.49\textwidth]{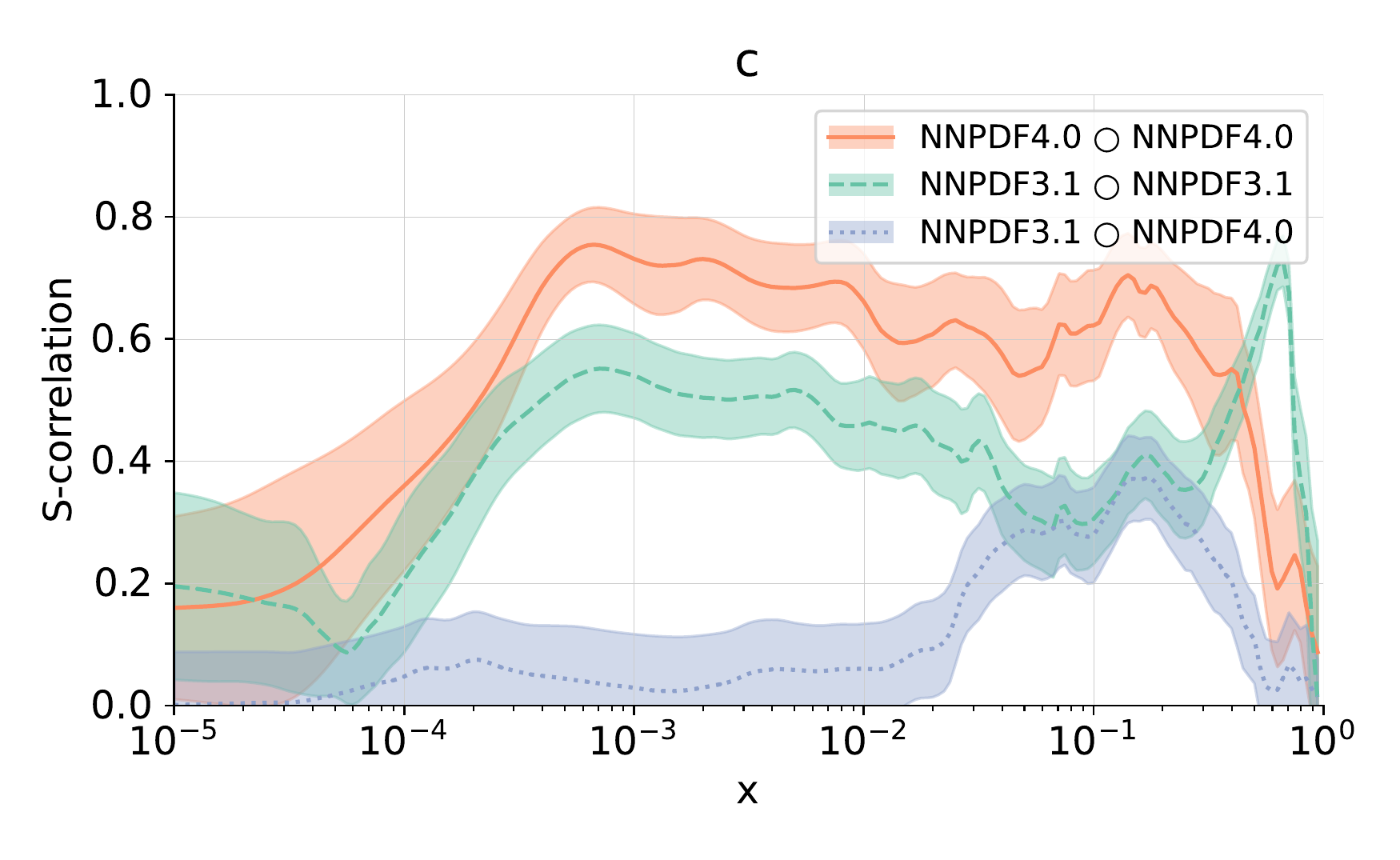}%
    \includegraphics[width=0.49\textwidth]{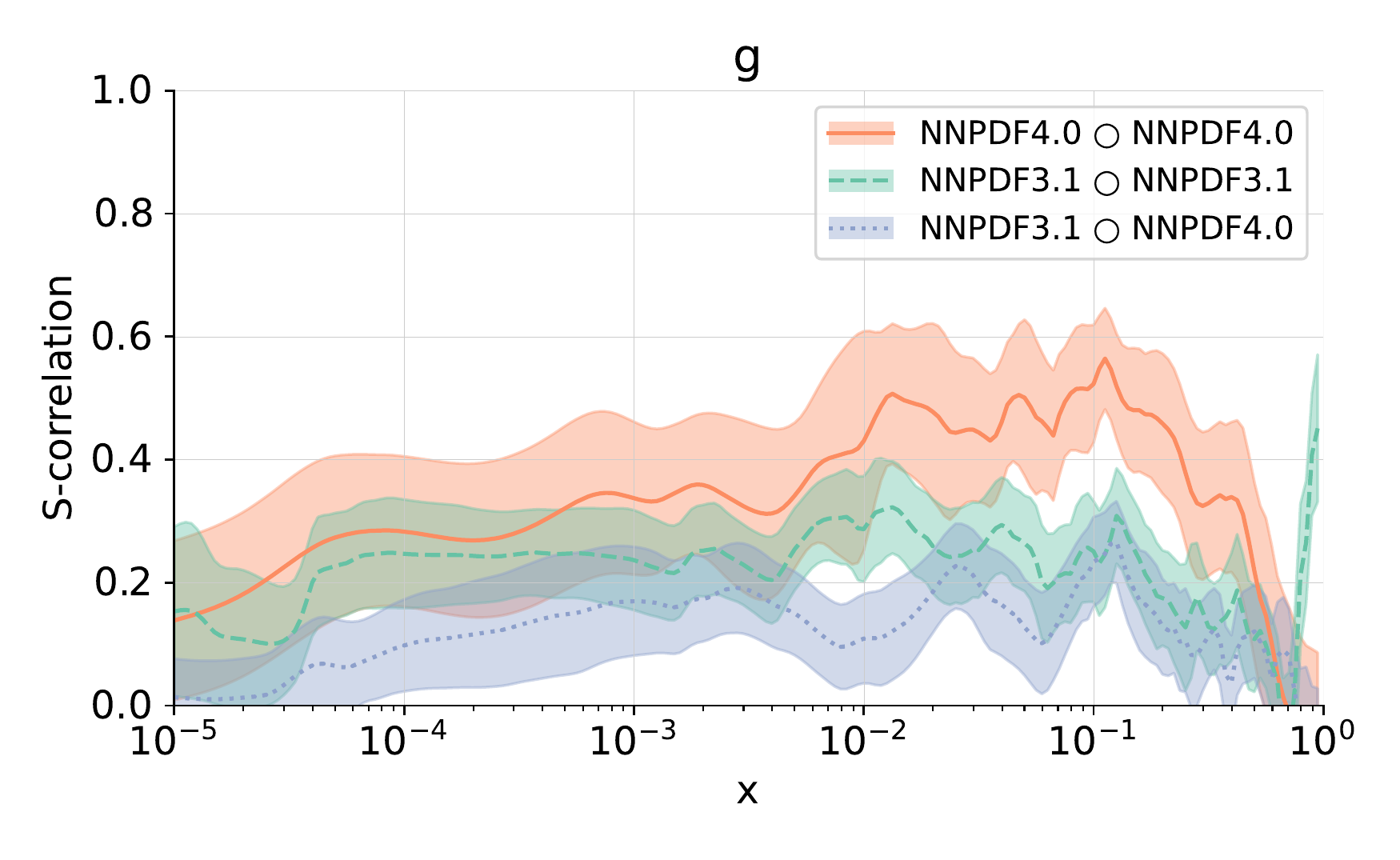}\\
    \caption{\small The data-driven component of the S-correlation
        Eq.~(\ref{eq:xcov}). Results are shown for all PDF
        flavors and the gluon. We consider PDF sets determined with
        the NNPDF3.1 methodology or the NNPDF4.0 methodology, and the
        three curves shown correspond to comparing pairs of sets with
        either methodology to themselves (NNPDF3.1: green; NNPDF4.0:
        orange) or with each other (blue). The shaded band for each curve
        is the 2$\sigma$ uncertainty due to the finite size of the replica sample,
        estimated by bootstrapping (see Appendix~\ref{app:correp}). Here
        and below $\bigcirc$ denotes the operation of comparing
        quantities computed from two sets of replicas that are correlated by
        being fitted to the same underlying data.
      \label{fig:selfcorr}}
  \end{center}
  \end{figure}

Having verified that samples of PDF replicas behave as expected, we
now proceed to the computation the data-induced component of the
S-correlation Eq.~(\ref{eq:selfcorrpdfdiag}), as discussed in Sect.~\ref{sec:correp}.
Results are shown in
Fig.~\ref{fig:selfcorr}, where we show the S-correlation between
two sets of replicas determined with the NNPDF3.1 methodology (orange);
between
two sets of replicas determined with the NNPDF4.0 methodology (green);
and between  a set of replicas determined with the NNPDF3.1 methodology and
a set of replicas determined with the NNPDF4.0 methodology (blue). The error bands
show the 2$\sigma$ uncertainty due to the finite size of the replica
set, estimated using bootstrapping (see Appendix~\ref{app:correp}).

It is apparent from the plots that the data-induced PDF S-correlation drops
very quickly to zero outside the data region, as it ought to. For light
quarks, the correlations drop to zero for $x\lesssim 10^{-4}$ and $x\gtrsim 0.4$;
while the data region is rather smaller for the the gluon and heavy quarks.
This is because in a DIS-only
PDF determination the gluon PDF is only
determined indirectly by scaling violations, while the PDFs for heavier quarks
are determined by charged current data, i.e. mostly by
fixed-target neutrino DIS data. However, interestingly, even in the
middle of the data region the S-correlation of pairs of PDF sets determined
using the NNPDF3.1 methodology is typically around
40\% and never exceeds 60\%. The S-correlation of PDF sets determined
using the NNPDF4.0 methodology, in turn, is typically around
60\% and never exceeds 80\%. The S-correlation between PDF sets
determined using the NNPDF3.1 and NNPDF4.0 methodologies, finally, is very
similar to the S-correlation between the pair of NNPDF3.1 PDF sets.

As discussed in Sect.~\ref{sec:correp}, the deviation from unity of
the data-driven S-correlation when comparing of a PDF set to itself is a measure of the size
of the functional component of the S-correlation. Both for the
NNPDF3.1 and NNPDF4.0 methodologies, this deviation is substantial.
 Note that, as shown in
Fig.~\ref{fig:distances}, any two independent sets of replicas for the
same set have the same mean
and uncertainty within finite-size fluctuations, and that these
fluctuations scale as expected and in particular go to zero in the
limit of a large number replicas.
Note also that the deviation of the S-correlation from 100\% is
much larger than the uncertainty due to the finite size of the replica
sample, so the correlation loss cannot just be due to an insufficient
number of replicas.

The fact that the  S-correlation of NNPDF3.1 PDFs is smaller than
that of NNPDF4.0 PDFs is consistent with the
fact~\cite{Forte:2020yip,Cruz-Martinez:2021rgy,Ball:2021leu}
that the NNPDF4.0 methodology leads to smaller uncertainties than the
NNPDF3.1 methodology, even though both can be shown to be faithful using closure
tests~\cite{Ball:2014uwa,Ball:2021leu}. Indeed, the only way  PDF sets
determined from the same underlying data can have different uncertainties
is if one of the two has a smaller functional (i.e. non-data-driven)
component of the uncertainty. But then we
would  expect that the methodology characterized by a smaller
functional uncertainty also has a smaller functional S-correlation:
i.e. that it is determined to a greater extent by the underlying
data. This is indeed what happens here: NNPDF4.0 has a smaller
uncertainty for a fixed dataset, and accordingly a larger S-correlation.

  \clearpage

It is interesting to observe that  the data-induced component of the
S-correlation between 
NNPDF3.1 and NNPDF4.0 PDFs is almost always similar to that of the PDF set
that has the smallest correlation to itself, namely NNPDF3.1 (a possible
exception being the charm PDF, which is a special case because in a
DIS-only fit it is almost undetermined). This
suggests a ``weakest-link'' explanation: the  data are  more weakly
correlated to NNPDF3.1 than to NNPDF4.0,
and so inevitably the data-driven correlation
between NNPDF3.1 and NNPDF4.0 is dominated by this weaker correlation.
This is apparent, for instance,  in the small $x$ region, where the
data-driven S-correlation of NNPDF3.1 to itself is significantly
weaker than that of NNPDF4.0.

All this suggests that the data-driven component of the  S-correlation between PDF sets
for a given methodology can be used as a criterion for
the assessment of the efficiency of the methodology itself, with the
interpretation that a methodology leading to higher cross-correlation is more
efficient.  Namely, PDFs determined using a  methodology characterized
by higher  S-correlation have a smaller functional component of
the S-correlation, i.e. they are to a greater extent determined by the
underlying data.
So for instance the weaker S-correlation of NNPDF3.1 at
small $x$ suggests that in this region the NNPDF3.1 uncertainties
could be reduced without loss of accuracy, as is indeed the case~\cite{Ball:2021leu}.

In order to further investigate the functional component of the
S-correlation, we have produced  sets of PDF replicas in
which some methodological choices are correlated or decorrelated.
First, we have produced a set of replicas in which preprocessing is
also correlated. To understand this, recall that neural networks used to
parametrize PDFs include a preprocessing function~\cite{Forte:2020yip},
whose parameters are randomly varied between replicas. We have thus
produced a new pair of NNPDF4.0 replicas in which not only the data,
but also the preprocessing exponents are correlated: so in
Eq.~(\ref{eq:xcov}), replicas ${f_{a}^p(x,Q_0^2)}^{(r)}$ and
${f_{b}^p(x,Q_0^2)}^{(r)}$ are not only fitted to the same underlying
    data, but also have the same value of the preprocessing
    exponents.

\begin{figure}[t]
\begin{center}
  \includegraphics[width=0.49\textwidth]{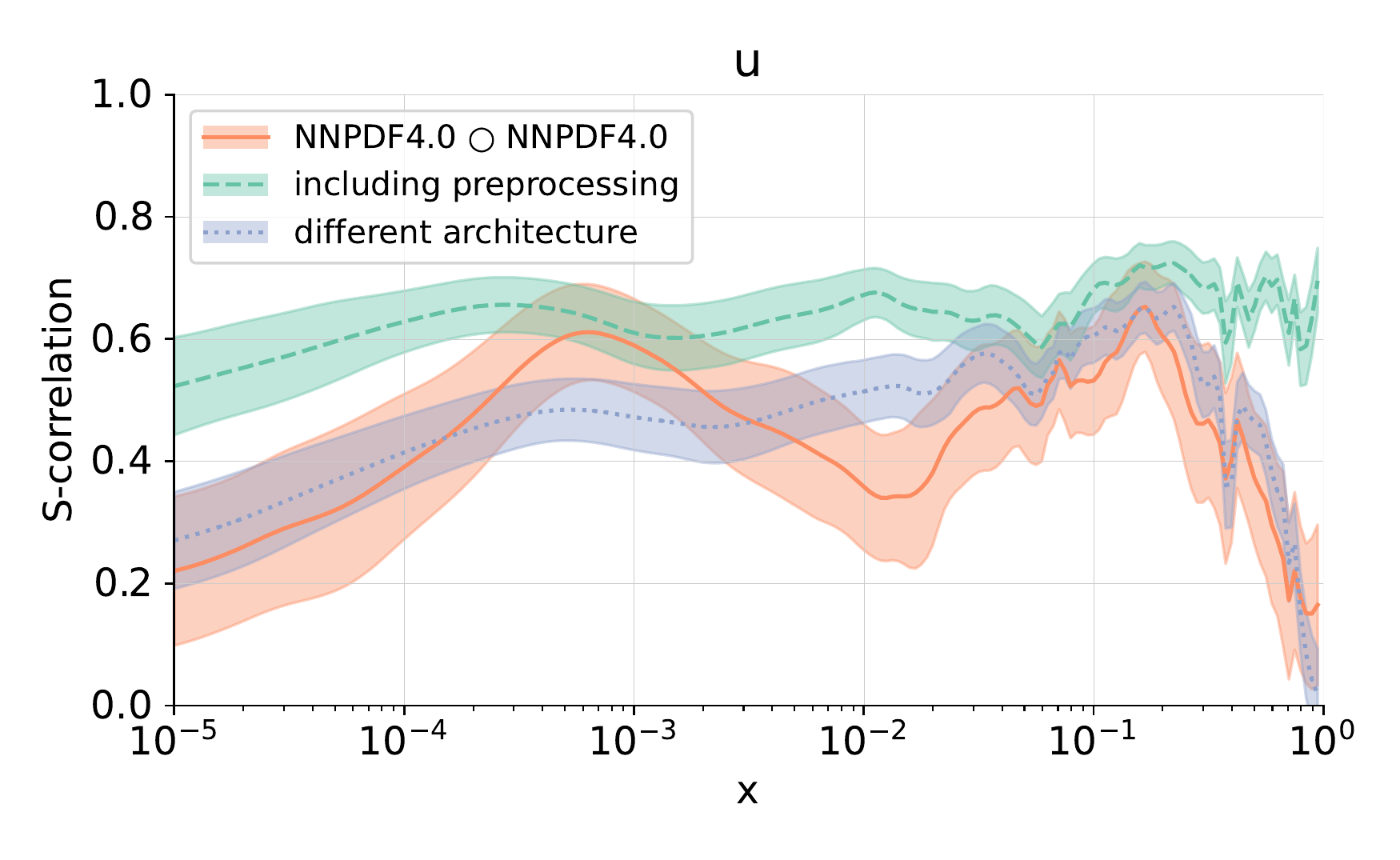}%
  \includegraphics[width=0.49\textwidth]{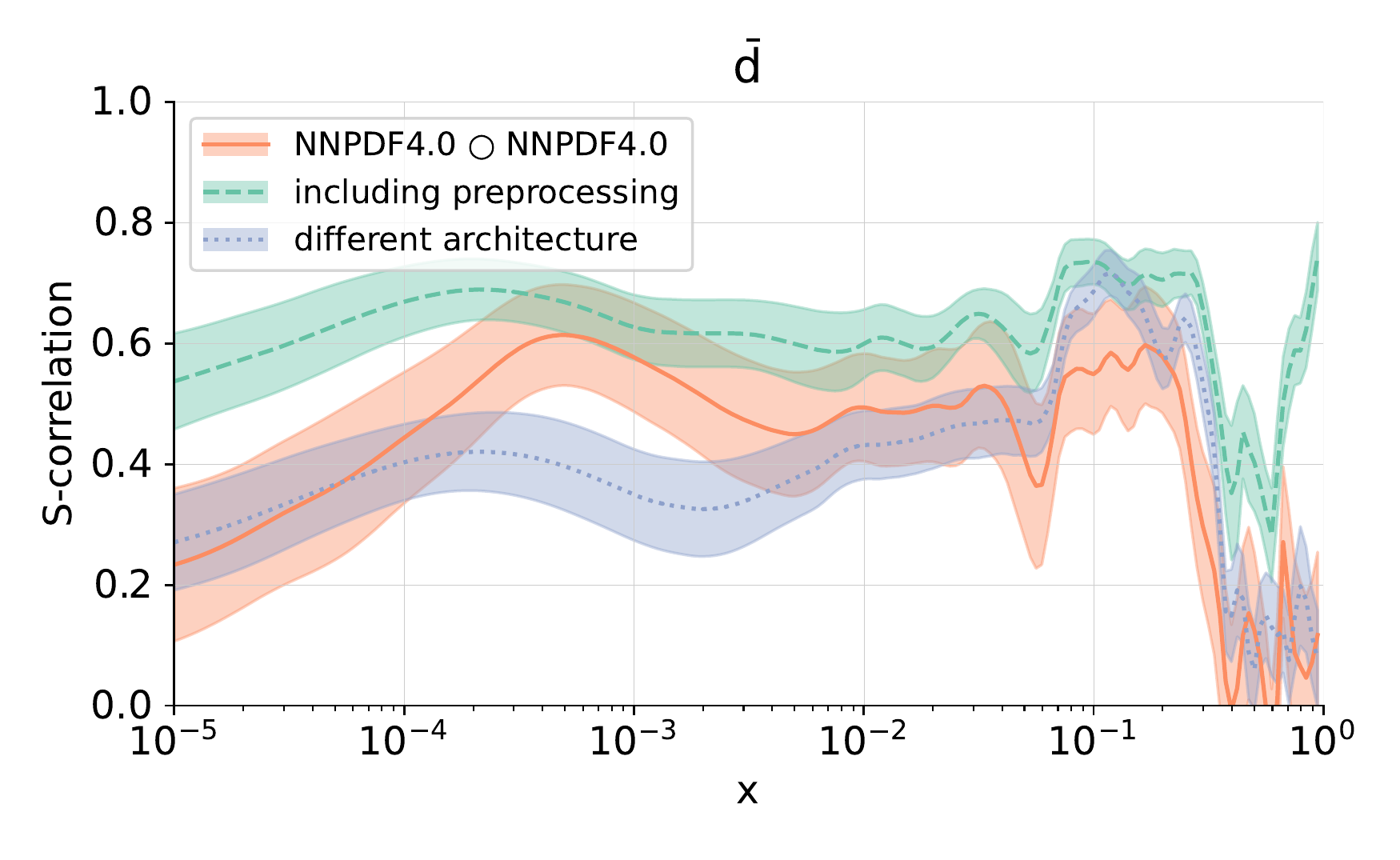}\\
  \includegraphics[width=0.49\textwidth]{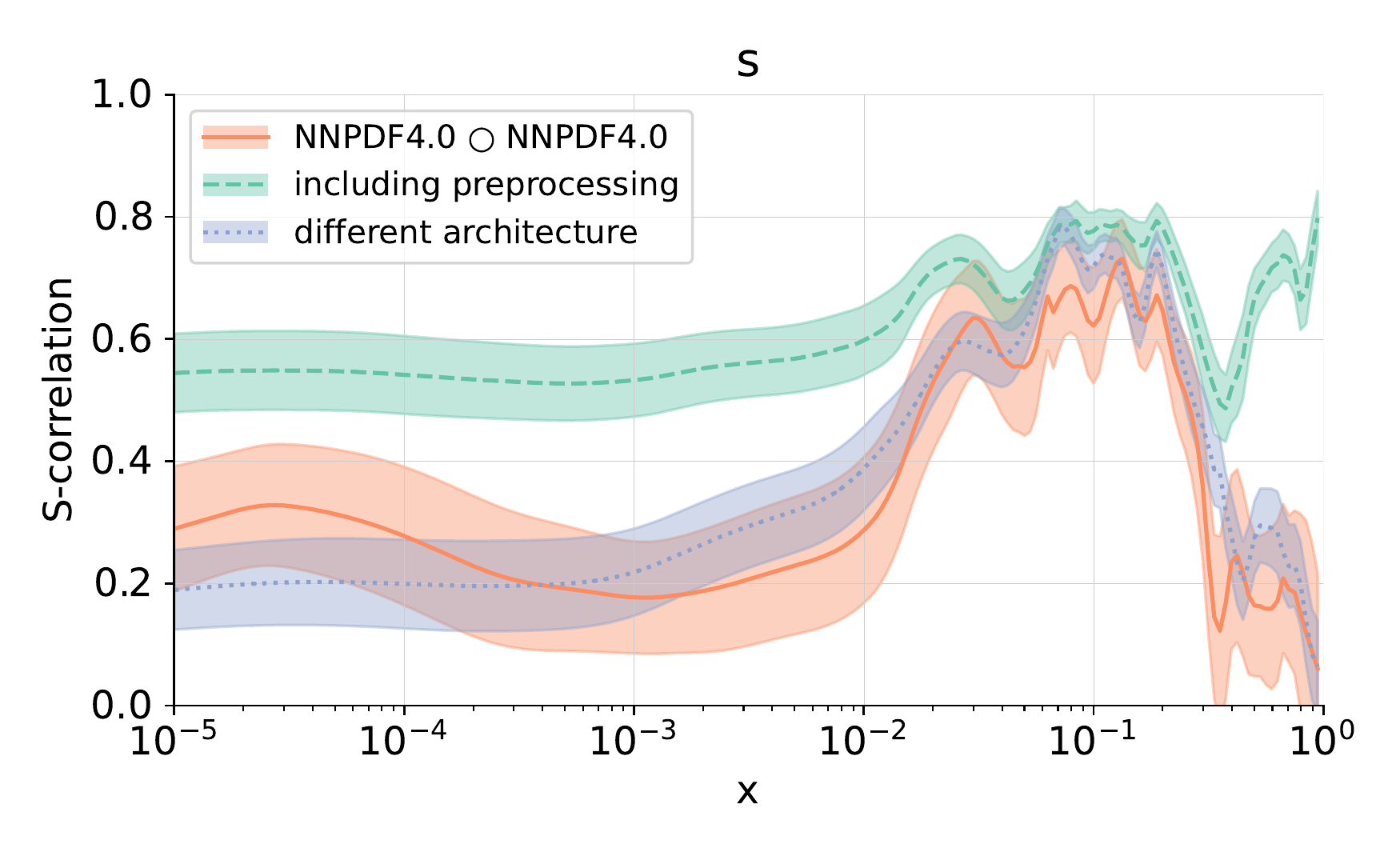}
  \includegraphics[width=0.49\textwidth]{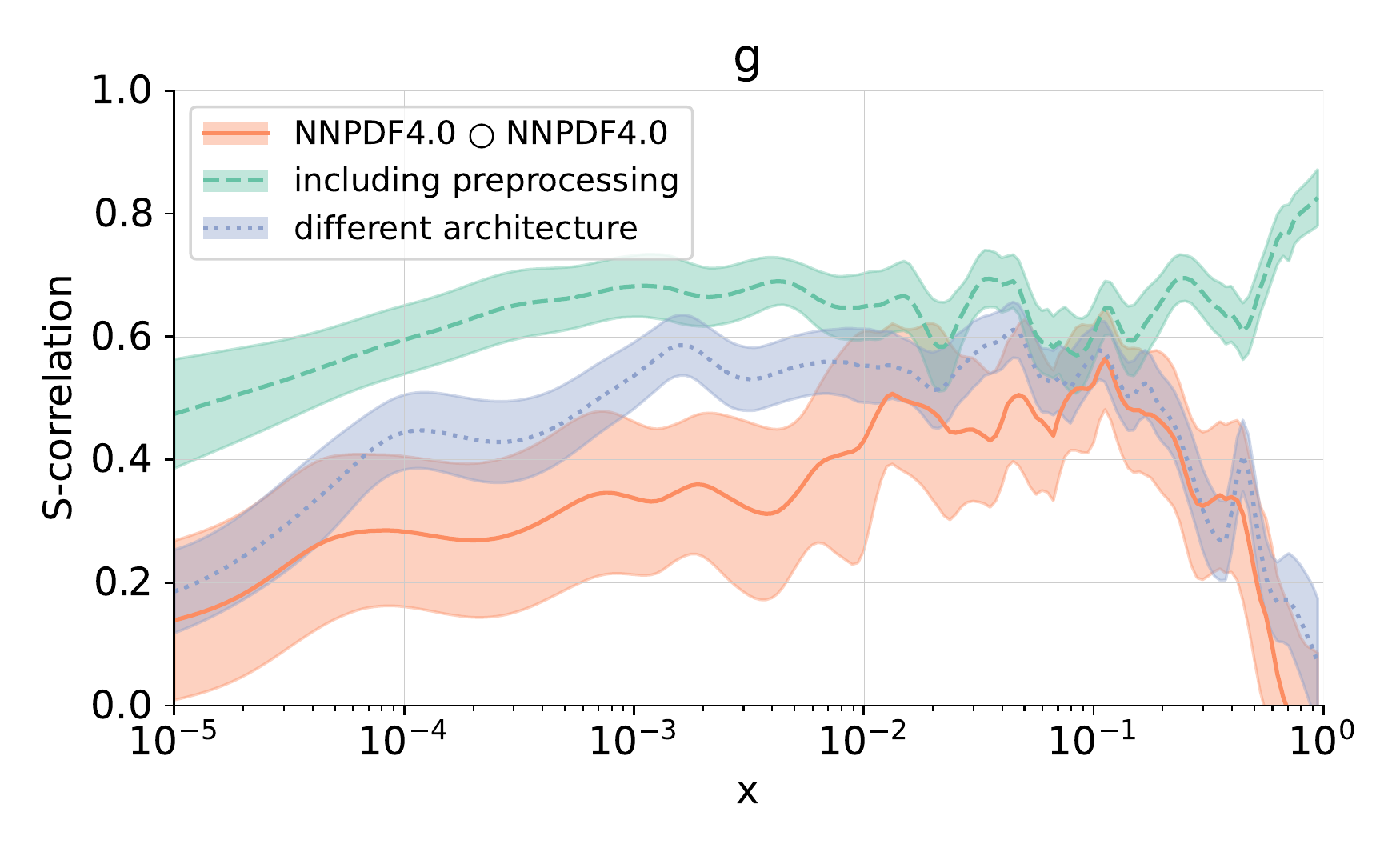}%
  \caption{\small The data-driven component of the S-correlation
      Eq.~(\ref{eq:xcov}) between PDFs determined with the NNPDF4.0
      methodology (same as
      Fig.~\ref{fig:selfcorr}) compared to the case in which also
  the preprocessing-induced  component is included in the
  correlation (green), and the case in which the neural network architecture
  is changed with all other aspects of the methodology kept fixed (blue).
  Results are shown for the up, anti-down, strange and gluon PDFs.
    \label{fig:selfcorrvar}}
\end{center}
\end{figure}

Results are shown in green in Fig.~\ref{fig:selfcorrvar}, compared to the
previous results of Fig.~\ref{fig:selfcorr}, shown in
orange.
It is clear that in the data region, where the data-induced S-correlation is
largest, the extra correlation due to preprocessing is negligible. As
PDFs extrapolate further away from the data region, the
contribution due to preprocessing is increasingly large: for instance
for the gluon at large $x\gtrsim0.4$ the data-induced correlation
rapidly drops to zero as $x\to1$, but the correlation due
preprocessing makes up for the decrease and in fact it somewhat
exceeds it as the kinematic boundary at $x=1$ is approached.

Furthermore, we have produced a PDF set, based on the NNPDF4.0 methodology,
but with a different architecture of the neural net, i.e., different number of layers and
layer sizes. This is thus
effectively a variation of the NNPDF4.0 methodology. 
Results, also shown in
Fig.~\ref{fig:selfcorrvar} (in blue), demonstrate that this specific
aspect of the methodology has little impact: the S-correlation is
essentially the same as in the case where the
architecture of the neural networks is the same in the two sets being
compared. This shows that these two methodologies lead to very similar
results, which in turn suggests that correlating the neural network
architecture would have a less significant impact than that of
correlating preprocessing.

These two examples illustrate how, at least in principle, all
components of the S-correlation
could be determined, namely, by correlating all methodological aspects
that determine the final result. As already discussed, 
whereas this is easily done for parametric choices (like the values of
the preprocessing exponents), it is rather more difficult
for non-parametric aspects, such as, for instance, the choice of
minimization settings. These aspects are of course  closely tied to the non-uniqueness
of the best fit for given data, which leads to functional uncertainties.

\begin{figure}[t]
\begin{center}
  \includegraphics[width=0.49\textwidth]{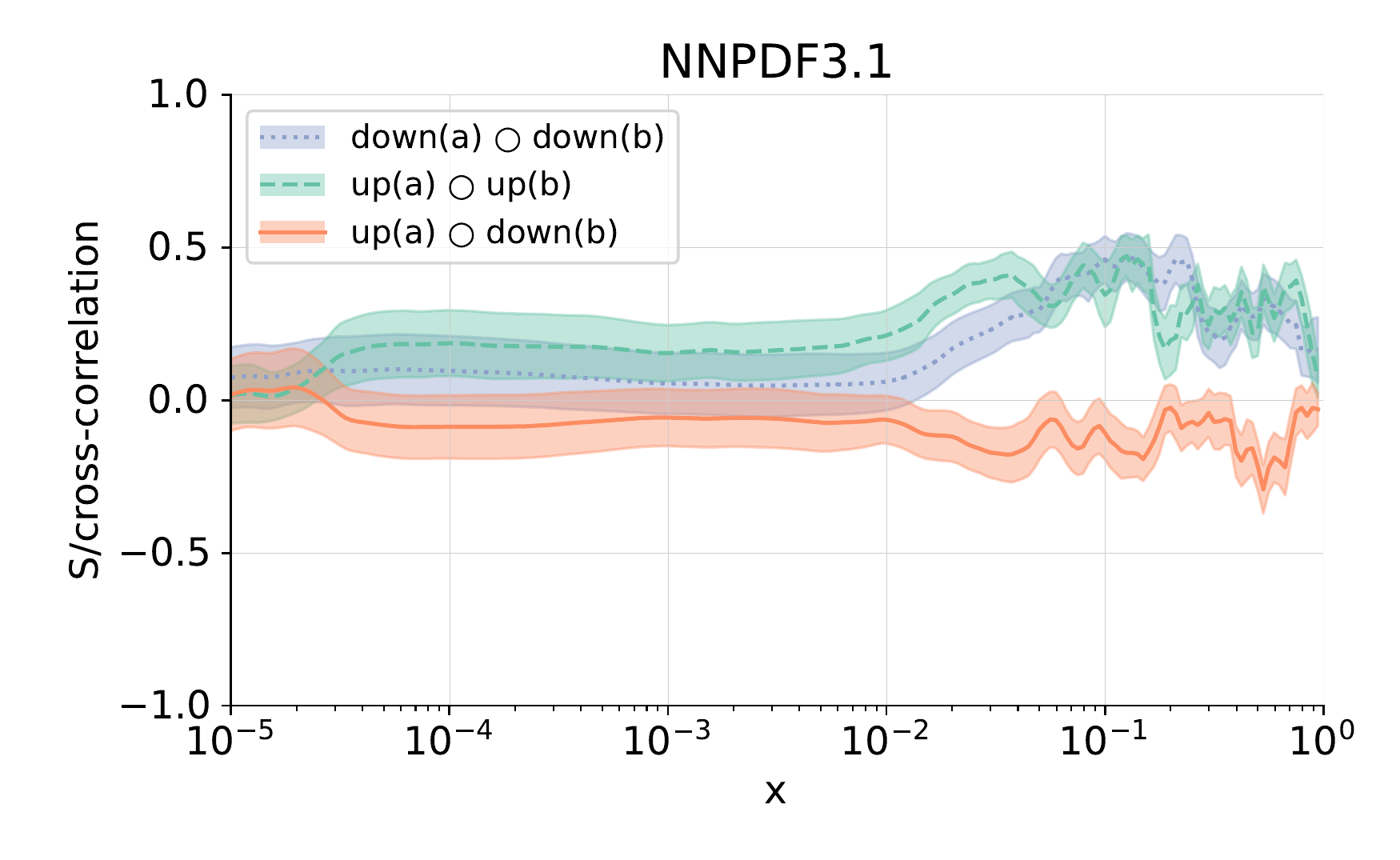}%
  \includegraphics[width=0.49\textwidth]{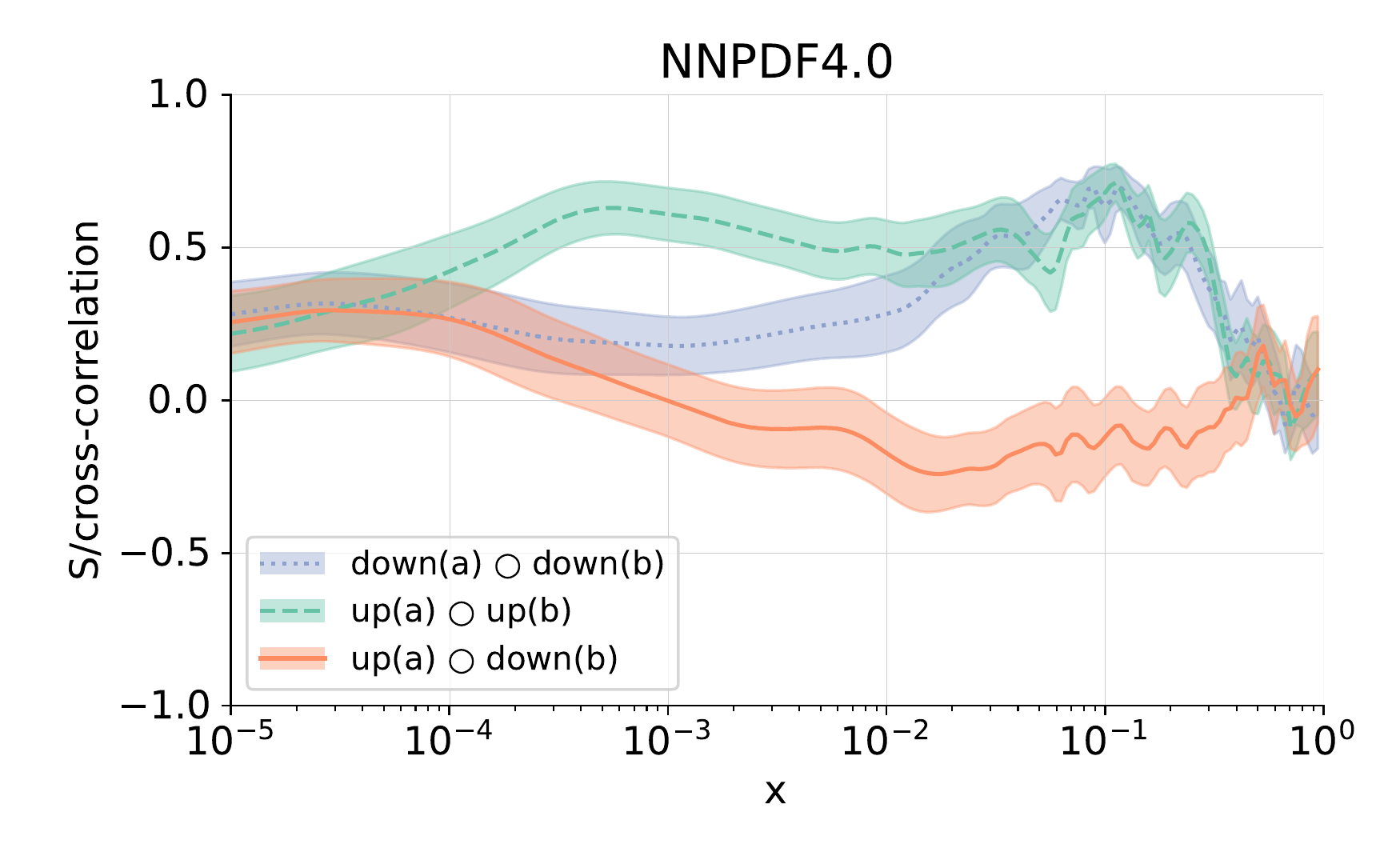}\\
  \includegraphics[width=0.49\textwidth]{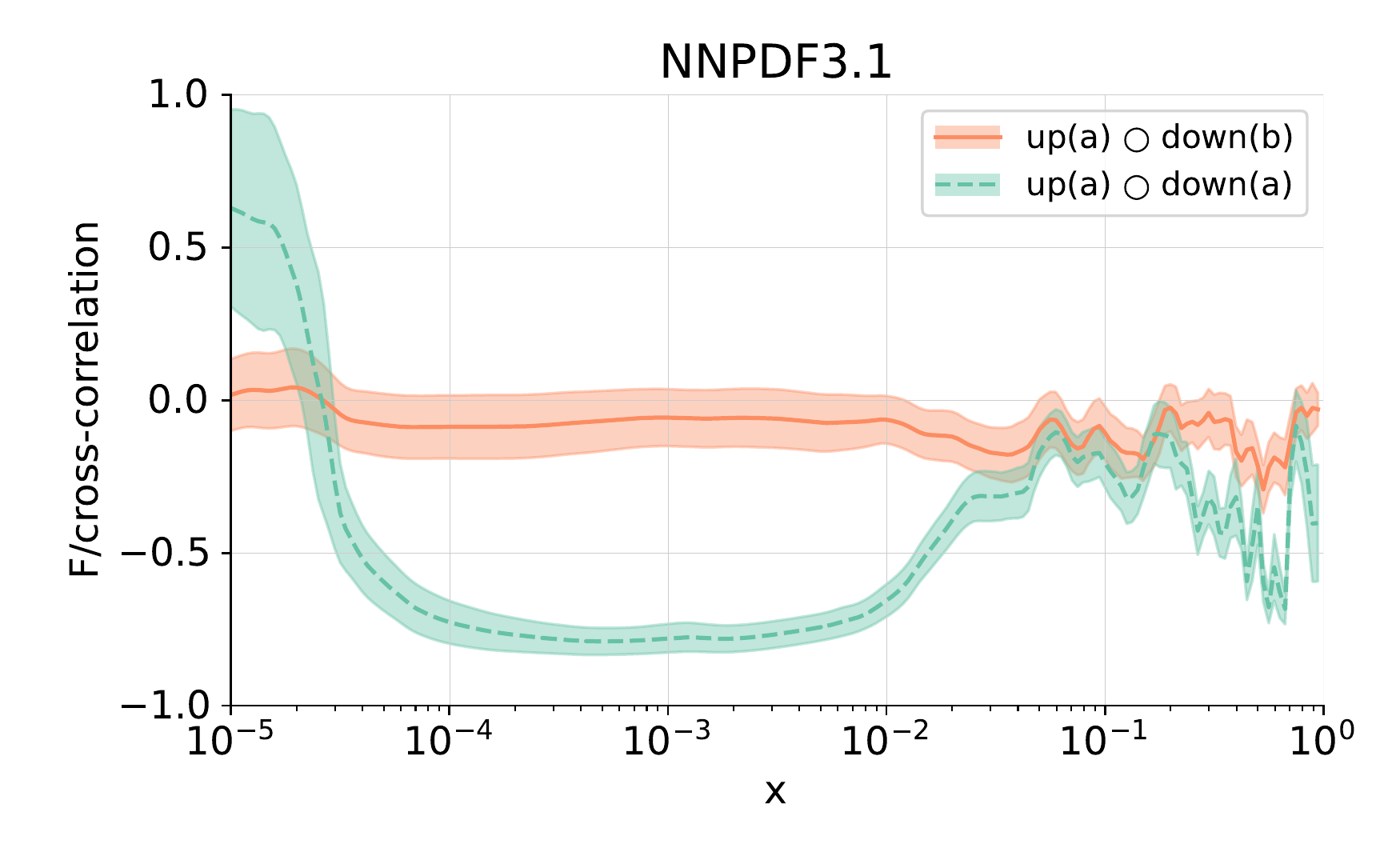}
  \includegraphics[width=0.49\textwidth]{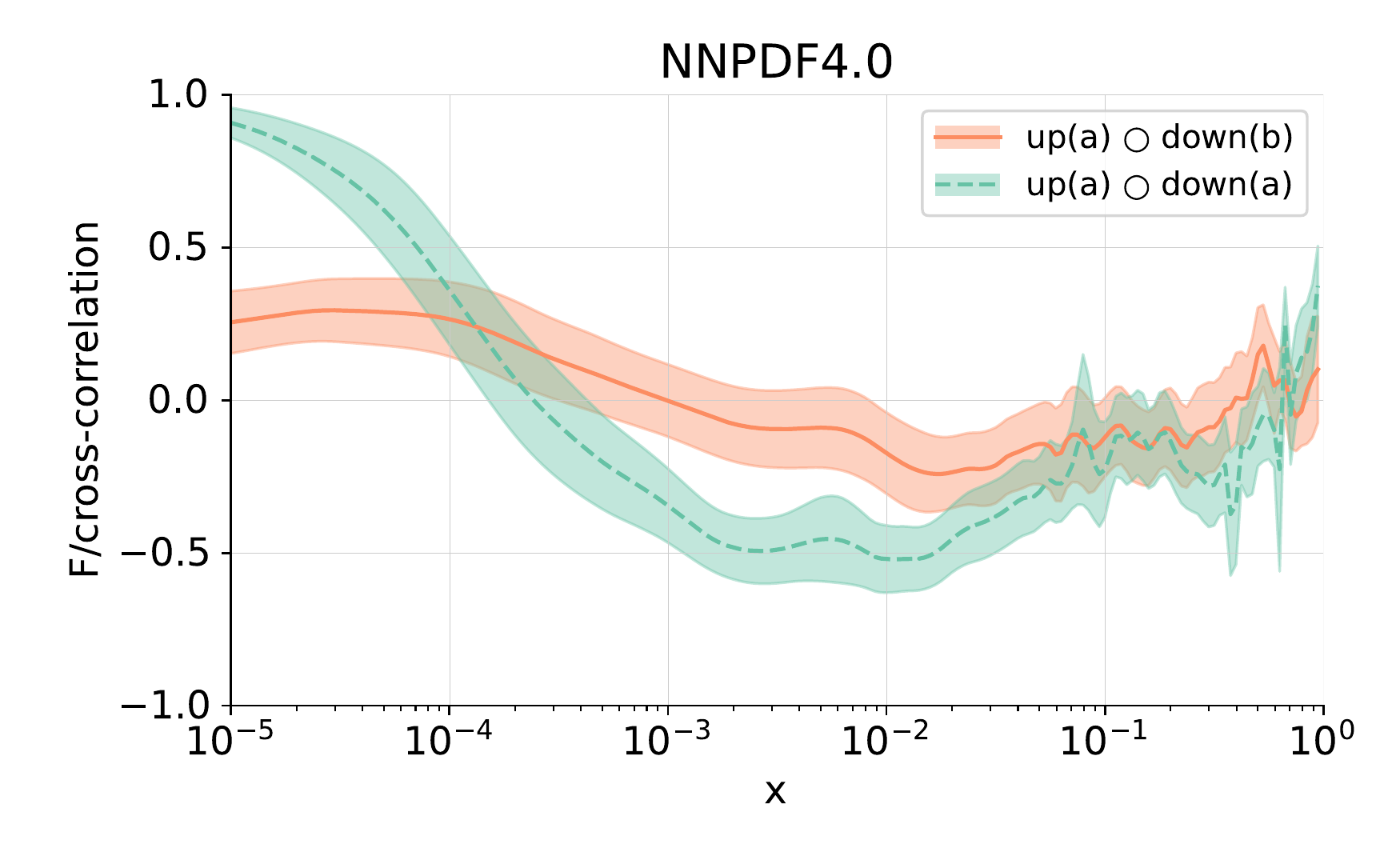}%
  \caption{\small Top: comparison of the the data-driven component of
    the S-correlation Eq.~(\ref{eq:xcov})
    of the up and down PDFs and the up-down  cross-correlation
      Eq.~(\ref{eq:selfcorrpdf}).
    Bottom: comparison between the data-driven component of the up-down
 cross-correlation Eq.~(\ref{eq:selfcorrpdf}) and the standard
      F-correlation Eq.~(\ref{eq:corrpdf}) for the up and down quark
      PDFs. Results are shown for the NNPDF3.1 (left) and the NNPDF4.0
      methodology (right). In all plots (a) and (b) denote two
      distinct sets of
      correlated replicas (i.e. fitted to the same underlying data),
      so (a)~$\bigcirc$~(b) denotes the case in which the replicas are
      correlated but distinct while (a)~$\bigcirc$~(a)  denotes the
      case in which the replicas are identical.   
    \label{fig:selfcorrud}}
\end{center}
\end{figure}

Finally, we compare, for a fixed methodology, the data-driven
component of the S-correlation for a pair of PDF flavors,  to
the cross-correlation  between them:
in Fig.~\ref{fig:selfcorrud} results are shown for the up and the
down PDFs,
both for the NNPDF3.1 methodology and the NNPDF4.0 methodology. Clearly,
unlike the diagonal S-correlation in the flavor basis
Eq.~\eqref{eq:selfcorrpdfdiag}, the correlation between two different
PDF flavors need not be positive. This is indeed seen in
the figure, namely, the up and down PDFs turn out to be anticorrelated at large
$x$. Furthermore, one would generally expect any  cross-correlation between two
different PDF flavors to be weaker than the S-correlation --- indeed, if
all sources of S-correlation were included, the S-correlation
would be 100\%. This is again borne out by the explicit computation,
that shows that the  cross-correlation is generally smaller in
modulus than the S-correlation. Note that for the less efficient
NNPDF3.1 methodology, for which all S-correlations are smaller, as
already discussed, the cross-correlation is accordingly smaller
in modulus.

In the same figure we also compare, again for a fixed methodology,
the data-induced component of the
cross-correlation between
the up and down PDFs to the standard F-correlation
Eq.~\eqref{eq:corrpdf}. 
Note that if the cross-correlation was entirely
data-driven, for a fixed methodology these two quantities would
coincide. But in actual fact they differ because in the computation of
the F-correlation, Eq.~(\ref{eq:corrav}) is used (the replicas are
fully correlated), while in the computation of the cross-correlation,
Eq.~(\ref{eq:scorrav}) is used (the replicas are only correlated
through data).
Whenever the S-correlation is
sizable, the data-induced
component of the cross-correlation, and the  F-correlation are very
close to each other. This means that the functional component of the
PDF uncertainty is essentially uncorrelated between different PDFs,
i.e. that the standard  F-correlation is due to the correlation
between the underlying data. This is as one expects, and
justifies using PDF correlations to estimate
the impact of data uncertainties on PDFs uncertainties  and
conversely.

However, when the data-induced component of the S-correlation is small,
then the S-correlation can differ significantly from the 
F-correlation.
This is clearly seen when comparing
the up-down correlation in the region $10^{-4}\lesssim x \lesssim
10^{-2}$ computed with the NNPDF3.1 methodology to that computed with the NNPDF4.0 methodology. With the
NNPDF4.0 methodology, the data-driven S-correlation in this region is large,
and the standard up-down correlation is quite close to the 
cross-correlation. With the NNPDF3.1 methodology, on the other hand, the data-driven
S-correlation is
almost vanishing. This means that, with NNPDF3.1  methodology, PDFs in this
region are completely dominated by functional uncertainty, which then
screen out the PDF correlation when computing the 
cross-correlation. In other words, the functional component of the
S-correlation (between a PDF and itself) is generally non-negligible, while the functional
component of the  cross-correlation (between two different
PDFs or two different $x$ values) is generally quite small.

\section{Combined PDF sets}
\label{sec:comb}

As already mentioned in Sect.~\ref{sec:intro}, combined PDF sets have been
produced~\cite{Botje:2011sn,Butterworth:2015oua}, with the goal of
providing a common, conservative PDF determination. The underlying
idea is that so-called ``global'' PDF sets, namely PDFs determined
using the widest possible amount of experimental information available
at a given time, differ due to theoretical and methodological
assumptions. If these assumptions all satisfy reasonable criteria of
reliability,
these different PDF determinations are considered to be
equally likely, and thus a conservative choice is to combine them into
a single determination. One may then reasonably ask whether this
combination might be constructed in such a way as to explicitly take 
account of the cross-correlation between PDF sets.

\subsection{The PDF4LHC15 combination}
\label{sec:stat}

\def\smallfrac#1#2{\hbox{$\frac{#1}{#2}$}}
\def\half{\smallfrac{1}{2}}
\def\third{\smallfrac{1}{3}}
\def\quarter{\smallfrac{1}{4}}
\def\nn{\nonumber}

The PDF4LHC15 prescription~\cite{Butterworth:2015oua}
for combining different PDF sets $\Phi_a,\,\Phi_b,\dots$ may be described as
follows. All sets being combined are turned into a Monte Carlo
representation, as discussed in Sect.~\ref{sec:correp}, i.e. each of them
is represented by a set of $N$ PDF replicas $\{{f_i^p}{(r)}: r =
1,\ldots N\}$, where the index $i$ runs over the sets that are being combined.
The combined set is then defined simply as the union of the replicas in the
individual sets: specifically to combine two sets $\Phi_a$ and $\Phi_b$, we
select randomly $\half N$ replicas from each set, and then define the replicas
for the combined set $\{F^p\} = \{{F^p}^{(r)}: r = 1,\ldots N\}$ as
\begin{equation}
{F^p}^{(r)} = \begin{cases}{f_a^{p(r)} \qquad\mbox{for $r = 1,\ldots,\half N$;}}\\{ f_b^{p(r)}\qquad \mbox{for $r = \half N+1,\ldots,N$.}}\end{cases}
\label{eq:Fdef}
\end{equation}
The combination assumes
that, in the absence of an objective criterion for deciding on the relative
probability of the different sets, they are  equally probable, and thus that we
should take the same number of replicas from each set.

The combined set $\{F^p\}$ is then treated in the same way as the individual sets
for the calculation of estimators, with the averages taken over all the replicas in
the set according to Eq.~(\ref{eq:EXf}). Thus in particular the mean
of any PDF in the
combined set is immediately seen to be given by
\begin{equation}
E[ F^p] =\half ( \langle f^p_{a}\rangle + \langle f^p_{b}\rangle),
\label{eq:EF}
\end{equation}
while the F-covariance between two PDFs is
\begin{eqnarray}
{\rm Cov}[F^p,F^q] &=& \langle  F^p F^q\rangle -  \langle F^p\rangle\langle F^q\rangle =
\half( \langle f^p_{a}f^q_{a}\rangle + \langle f_{b}^pf_{b}^q\rangle)  -  \quarter
( \langle f_{a}^p\rangle + \langle f_{b}^p\rangle)( \langle f_{a}^q\rangle + \langle
f^q_{b}\rangle)\nn \\
&=& \half( {\rm Cov}[f^p_{a},f^q_{a}] + {\rm Cov}[f^p_{b},f^q_{b}]) +  \quarter ( \langle
f^p_{a}\rangle - \langle f^p_{b}\rangle)( \langle f^q_{a}\rangle - \langle
f^q_{b}\rangle).
\label{eq:CovF}
\end{eqnarray}
Thus in the combined set, the central PDF is the mean of the central PDFs in
each set (up to the usual statistical uncertainties of order $1/\sqrt{N}$),
whereas the uncertainties are always greater than the mean of the
uncertainties, the extra term being due to the spread of the central
predictions. This is as it should be: when the sets used in the
combination disagree, the uncertainty is increased.

The expressions Eqs.~(\ref{eq:EF},\ref{eq:CovF}) can be generalized
straightforwardly to the combination of $n$ PDF sets, by taking $N/n$ replicas
at random from each (always keeping $N\gg n$ of course): then
\begin{equation}
F^{p(r)} = \begin{cases}{f_{a_1}^{p(r)} \qquad\mbox{for $r = 1,\ldots,\smallfrac{N}{n}$;}}\\
{ f_{a_2}^{p(r)}\qquad \mbox{for $r = \smallfrac{N}{n}+1,\ldots,2\smallfrac{N}{n}$,}}\\
{\vdots}\qquad\qquad\qquad {\vdots}\\
{ f_{a_n}^{p(r)}\qquad \mbox{for $r = \smallfrac{n-1}{n}N+1,\ldots,N$,}}\end{cases}
\label{eq:Fdefn}
\end{equation}
and
\begin{equation}
E[F^p] = \smallfrac{1}{n} \sum_a \langle f^p_{a}\rangle, \qquad {\rm Cov}[F^p,F^q] =
\smallfrac{1}{n}\sum_a {\rm Cov}[f^p_{a},f^q_{a}] +  \smallfrac{1}{n^2} \sum_{a\neq b}(
\langle f^p_{a}\rangle - \langle f^p_{b}\rangle)( \langle f^q_{a}\rangle - \langle
f^q_{b}\rangle).
\label{eq:ECovFn}
\end{equation}
For large $n$ the extra term in the covariance of the combination increases the
result according to the covariance of the distribution of central values, since
in the pairwise sum there are $n(n-1)$ terms.

The PDF4LHC15 prescription is based on the assumption that the PDF sets
that are being combined, viewed as measurements of the true underlying
PDF, are all equally likely, which means that they have
approximately the same uncertainty and are approximately 100\%
correlated, i.e. they are not independent.
Indeed, independent (uncorrelated or partly correlated)
measurements of the same quantity a priori bring in new information on
the true value, so the  uncertainty on their combination
is always smaller or equal to the uncertainty of any of the measurements
that are being combined, (see e.g. Ref.~\cite{Cowan:1998ji}), as we
shall discuss further in
Sect.~\ref{sec:unc} below.

In fact,  an important property of the PDF4LHC15 combination prescription is that if the
constituent PDF sets $\Phi_a$ are perfectly correlated, meaning that taking
averages over replicas of the different PDF sets all give the same result, the
combination also gives this result. Note that perfectly correlated sets will
still be distinct, in the sense that the replicas will not be the same: it is
only the averages over the full ensemble of replicas that are the same, in the
limit when the number of replicas $N$ becomes very large. An example
are the various replica sets considered in  Sect.~\ref{sec:dcorr}, all
based on the NNPDF4.0 methodology: both the two sets compared in
Fig.~\ref{fig:distances}, with replicas based on different underlying
data replicas, and those compared in Fig.~\ref{fig:selfcorr}, based on
correlated underlying data replicas. For these pairs of sets, $\langle
f^p_{a}\rangle = \langle f^p_{b}\rangle$ for all $a,b$, and  ${\rm
Cov}[f_a^p,f_a^q] = {\rm Cov}[f_b^p,f_b^q]$. This is of course true irrespective of the number
of sets $n$ used in the combination: it is just the same as when
combining several batches of PDF replicas from a given PDF set (such
as NNPDF4.0) into a single larger replica set.

To ensure that these assumptions are reasonable, that is,
the  PDF sets used in the PDF4LHC15 combination are highly correlated, and as such can
reasonably be combined by giving equal weight to each set, a number of criteria
was adopted~\cite{Butterworth:2015oua}:
\begin{itemize}
\item Each set is based on a global dataset, and in practice these global
datasets are very similar, both in size and content.
\item Theoretical calculations are performed to the same order in $\alpha_s$,
using a VFNS, and benchmarked against one another.
\item External parameters such as $\alpha_s$ and quark masses are given common
values where possible.
\item Each PDF determination includes procedural and functional uncertainties
in the adopted methodology.
\end{itemize}
Furthermore, an extensive benchmarking was performed in order to make
sure that indeed uncertainties from the various sets were
approximately equal, and that these criteria were  sufficient to ensure that the PDF sets used in the
combination could be be meaningfully assigned equal probability in
the combination.

\subsection{Correlated PDF combination}
\label{sec:unc}

It is clear that even though the PDF sets included in the PDF4LHC15
combination are highly correlated, the  correlation is manifestly not complete. Even assuming that the
benchmarking and parameter settings can achieve complete agreement, there will
still be some decorrelation through the choice of global dataset, and in the
different methodologies used by the different groups.
It has therefore been suggested \cite{froidevauxtalk:note} that a more
precise and accurate result might be obtained if different PDFs are
combined as independent, partly
correlated measurements of the underlying true PDF. The
 logic is that, even in the presence of a common underlying
 dataset, each PDF determination, based on a different methodology,
 might be extracting different information from
the data, just like different detectors could provide partly independent
though correlated information on the same physical phenomenon. A
correlated combination might then be advantageous because it would lead
to a more precise and accurate prediction.

Unbiased correlated measurements of the same underlying observable can
be combined in a standard way (see e.g.
Sect.~7.6 of Ref.~\cite{Cowan:1998ji}). Specifically, viewing the
expectation values $E[f_a^p]$ $a=1,2,\dots$ of PDFs as measurements of
an underlying true value, their correlated combination is a weighted
average, that we can in turn view as the expectation value of the
probability distribution for a combined determination $\tilde{F}^p$:
\begin{equation}
  E[\tilde{F}^p] = \sum_a w^p_a E[f_a^p],
  \label{eq:EFcowan}
\end{equation}
with  weights $w^p_a$  given by
\begin{equation}
  w^p_a = \frac{\sum_b \mathrm{Cov}^{-1}[f_a^p, f_b^p]}{\sum_{c,d} \mathrm{Cov}^{-1}[f_c^p, f_d^p]},
  \label{eq:wacowan}
\end{equation}
where $\mathrm{Cov}^{-1}[f_a^p, f_b^p]$ is the matrix inverse of the 
S-covariance. The square uncertainty on the combination 
Eq.~(\ref{eq:EFcowan}) is the variance of the probability distribution
of $\tilde{F}^p$, given by
\begin{equation}
  \mathrm{Var}[\tilde{F}^p] = \sum_{a,b} w^p_aw_b^p \mathrm{Cov}[f_a^p, f_b^p].
  \label{eq:VarFcowan}
\end{equation}
Of course, all this relies on the assumption that the measurements are
unbiased, and that  the correlation between them, namely, the S-correlation
Eq.~(\ref{eq:selfcorrpdfdiag}), can be reliably computed.

Even so, the combination Eq.~(\ref{eq:EFcowan}) is subject to several  caveats. Specifically,
 the weights $ w^p_a$ Eq.~(\ref{eq:wacowan}) depend not only on $p$
 but also on $x$ 
because the S-correlation does (recall Eq.~(\ref{eq:selfcovmatpdf})), and  consequently, the combined PDF
 Eq.~(\ref{eq:EFcowan}) does not automatically satisfy sum
 rules. Furthermore, for the same reason,
 the result of the combination will generally  depend on the scale
 at which it is performed, because with $x$-dependent weights even if
 the PDF sets $f_a^p$, for each $a$ satisfy QCD evolution equations,
 the combination does not. Finally, in order to compute physical
 observables using the combined PDFs, knowledge of the diagonal
 uncertainty Eq.~(\ref{eq:VarFcowan}) is not sufficient: rather, the
 full cross-covariance matrix for all $p,q$ and all $x,y$ would be
 required. This could be done in principle by sampling the PDFs,
 computing their F-correlation, and then turning the result into a
 Hessian representation by using techniques similar to those of
 Ref.~\cite{Carrazza:2015aoa}. A way out of all these problems might be 
to perform the weighted combination
Eq.~(\ref{eq:EFcowan}) not at the level of PDFs, but rather at the
level of physical observables. However, this has the further disadvantage that
cross-covariances and weights would have to be re-computed for each new
observable.  

Be all that as it may, our goal here is not to investigate
the most efficient way to implement the weighted combination, but
rather, to explore the implications of performing a correlated weighted
PDF combination according to Eqs.~(\ref{eq:EFcowan}-\ref{eq:VarFcowan}). 
As discussed in
Sect.~\ref{sec:stat}, in practice the PDF sets 
in the PDF4LHC15 combination
have approximately equal uncertainties. When this is the case, the weights
Eq.~(\ref{eq:wacowan}) are all approximately equal, and constant 
(independent of $x$), 
and then all the aforementioned problems can be ignored.
Indeed, when we combine
two PDF sets $\Phi_a$ and $\Phi_b$ 
such that
$ {\rm Var}[f_{a}^p]= {\rm Var}[f_{b}^p]$ the S-covariance  is
\begin{equation}\label{eq:correqw}
  \mathrm{Cov}[f_a^p, f_b^p] = \big(\delta_{ab}+(1-\delta_{ab})\rho[f_a^p,f_b^p]\big)\mathrm{Var}[f_a^p],
\end{equation}
from which it follows, using Eq.~(\ref{eq:wacowan}), that $w_a=w_b=\half$.

This equal weight situation
can be very simply  implemented in a Monte Carlo approach, in a
completely equivalent way. Indeed, assuming
that Monte
Carlo replicas are available for two PDFs, $\Phi_a$ and $\Phi_b$, the
correlated combination is
found by combining the two sets of replicas  into a single replica set given by 
\begin{equation}
\tilde{F}^{p(r)} = \half({f_a^p}^{(r)}+ {f_b^p}^{(r)})
\label{eq:Fdefcor}
\end{equation}
for $r = 1,\ldots, N$. Then $E[\tilde{F}^p]=\langle F^{p(r)}\rangle = 
\half(\langle{f_a^p}^{(r)}\rangle+ \langle{f_b^p}^{(r)}\rangle)$, is in 
agreement
with Eq.~(\ref{eq:EFcowan}) when $w_a=w_b=\half$. 
The F-covariance  Eq.~(\ref{eq:CovF}) evaluated over the
replica set Eq.~(\ref{eq:Fdefcor}) is now given by
\begin{eqnarray}
{\rm Cov}[\tilde{F^p},\tilde{F^q}] &=& \langle \tilde{F^p}\tilde{F^q}\rangle -  \langle \tilde{F^p}\rangle\langle \tilde{F^q}\rangle \nn\\
&=& \quarter( \langle f^p_{a }f^q_{a }\rangle + \langle f^p_{b}f^q_{b}\rangle + \langle (f^p_af_b^q+(f_b^pf_a^q)\rangle  -  \quarter ( \langle f_{a}^p\rangle + \langle f_{b}^p\rangle)( \langle f_{a}^q\rangle + \langle f^q_{b}\rangle)\nn \\
&=& \quarter( {\rm Cov}[f^p_{a},f^q_{a}] + {\rm Cov}[f^p_{b},f^q_{b}]) +  \quarter ({\rm Cov}[f^p_a,f^q_b]+{\rm Cov}[f^p_b,f^q_a]).
\label{eq:CovFcor}
\end{eqnarray}
Note that  when expressed in terms of the replicas from the
original sets $\Phi_a$ and $\Phi_b$ the F-covariance between two PDFs
in the combined set now depends on the cross-covariance between the
corresponding PDFs of the original sets ${\rm Cov}[f_a^p,f_b^q]$
Eq.~(\ref{eq:selfcorrpdf}).
Considering  the diagonal case $p=q$ in
Eq.~(\ref{eq:CovFcor}), the variance of the PDFs of the combined set
now depends on the S-correlation, and, using  $ {\rm Var}[f_{a}^p]=
{\rm Var}[f_{b}^p]$, Eq.~(\ref{eq:CovFcor}) with
$p=q$ reduces to
\begin{equation}
{\rm Var}[\tilde{F^p}]=\half\left(1+\rho[f^p_a,f^p_b]\right){\rm Var}[f^p_{a}],
\label{eq:CovFcorrd}
\end{equation}
which is the same as  Eq.~(\ref{eq:correqw}) when $a=b$.

Using  the correlated Monte Carlo approach, the properties of the
correlated combination are especially transparent.
Specifically,
it is clear   that the
uncertainty computed using Eq.~(\ref{eq:CovFcor}) is always smaller
than that found using the PDF4LHC15 combination.
To see this, note that the correlation  $|\rho [f_a,f_b]|\le 1$ or
equivalently $|{\rm Cov}[f_a,f_b]|\le \sqrt{{\rm Var}[f_{a}]}\sqrt{{\rm
      Var}[f_{b}]}$. It follows that the square uncertainty on
$F^p(x,Q_0^2)$ satisfies the inequality
\begin{equation}
{\rm Var}[\tilde{F^p}] \leq  \quarter({\rm Var}[f^p_{a}] + {\rm
  Var}[f^p_{b}])+  \half \sqrt{ {\rm Var}[f^p_{a}]} \sqrt{ {\rm
    Var}[f^p_{b}]}\leq   \half({\rm Var}[f^p_{a}]+{\rm Var}[f^p_{b}])\leq {\rm Var}[F^p],
\label{eq:doubineq}\end{equation}
where in going from the first to the second inequality  we
have trivially made use of the fact that $\half(x+y)^2\leq x^2+y^2$,
and the third inequality follows from the observation that the
second term in Eq.~(\ref{eq:CovF}) is non-negative.

The second inequality Eq.~(\ref{eq:doubineq}) has the  obvious implication that,
as already  mentioned in Sect.~\ref{sec:correp}, and as  seen
explicitly from Eq.~(\ref{eq:CovFcorrd}),  whenever the correlation
$\rho[f^p_a,f^p_b]<1$ the 
uncertainty of the correlated combination $\tilde{F}^p$ is smaller than either
of the uncertainties on $f_a^p$ or $f_b^p$, that have been assumed
to be approximately equal when forming the correlated combination
according to Eq.~(\ref{eq:Fdefcor}), ${\rm Var}[f^p_{a}]\approx{\rm
  Var}[f^p_{b}]$.

The third inequality Eq.~(\ref{eq:doubineq}) has the perhaps less obvious
implication   that  even if the two sets are fully correlated, so
$\rho[f^p_a,f^p_b]=1$,  the
uncertainty on the PDF4LHC15 combination  $F^p$, can be larger that the
uncertainty of either $f_a^p$ or $f_b^p$, even though
the uncertainty on the
correlated combination $\tilde{F^p}$ is  the same as that
 on both $f_a^p$ and $f_b^p$. This happens
whenever the central values of the two sets are not the same $\langle
f^p_a\rangle\not=\langle f^p_b\rangle$.

This is a
situation that the combination formula
Eq.~(\ref{eq:EFcowan})  cannot accommodate. Indeed, the uncertainty of
this
correlated combination can never exceed that
of the two determinations that are being combined.
This follows from the
assumption that the two determinations are unbiased estimators of
the same underlying true value.  Upon these assumptions, unit
correlation means that the
covariance matrix has a vanishing eigenvalue, so the
two determinations have the same
central value and uncertainty.

However, it is clearly
possible to have two random variables that have unit correlation but do
not have the same central value (or uncertainty). In particular the 
correlation of any two sets of random variables $f^r_1$ and $f^r_2$ is 
invariant under the linear transformations $f^r_1\to c_1f^r_1+k_1$, 
$f^r_2\to c_2f^r_2+k_2$, for any constants $c_1, c_2, k_1, k_2$, 
which change their mean values and variances. In the
Bayesian combination one simply takes the point of view that the two
measurements are equally likely determinations of the underlying true
quantity, so a priori they might be fully correlated, and yet  
their mean values and variances might differ. In such a situation, the 
variance of the PDF4LHC15
combination always comes out larger than those of the determinations that are
being combined. So, both the second and the third 
inequalities Eq.~(\ref{eq:doubineq})
become equalities only if PDF sets $\Phi_a$ and $\Phi_b$ are
identical, i.e. they have the same central value, uncertainty, and
unit correlation.

These results are easily generalized to the case of  $n$ PDF sets:
\begin{equation}
E[\tilde{F^p}] = \frac{1}{n} \sum_a \langle f^p_{a}\rangle, \qquad {\rm Cov}[\tilde{F^p},\tilde{F^q}] =  \frac{1}{n^2}\sum_a {\rm Cov}[f^p_{a},f^q_{a}] +  \frac{1}{n^2} \sum_{a\neq b}{\rm Cov}[f^p_a,f^q_b].
\label{eq:ECovFncor}
\end{equation}
In this case 
\begin{equation}
{\rm Var}[\tilde{F^p}] \leq  \frac{1}{n^2}\sum_a {\rm Var}[f^p_{a}] +  \frac{1}{n^2} \sum_{a\neq b}\sqrt{ {\rm Var}[f^p_{a}]} \sqrt {{\rm Var}[f^p_{b}]}\leq \frac{1}{n}\sum_a {\rm Var}[f^p_{a}] \leq {\rm Var}[F^p],
\end{equation}
and again equality can only be achieved when there is complete
equivalence between all the PDF sets in the combination, i.e. when
$f^{p(r)}_a=f^{p(r)}_b$ for all $r$, $p$ and for all pairs $a,b$. Otherwise,
the correlated combination inevitably reduces uncertainties. 

Even disregarding the issue related to PDF sets that have
different central values despite being very highly correlated, the
main  problem with the reduction in uncertainty in the correlated
combination
is that it is reliable
only if the cross-correlation has been correctly estimated. In
particular, if the cross-correlation is underestimated, then the
uncertainty on the combination is underestimated: in the most extreme
case, in which two PDFs are fully correlated, but the
cross-correlation is incorrectly determined to be very small,
the uncertainty on the combination is  underestimated by a factor
$\sqrt n$ (assuming again the
uncertainties of the $n$ starting PDFs are approximately the same).

In practice, the problem resides in the construction of  the correlated replicas
 to be used combination Eq.~(\ref{eq:Fdefcor}): this ought to
 be done in such a way that the averages over replicas in
 Eq.~(\ref{eq:CovFcor}) lead to a faithful determination of the
 F-covariance and the S-covariance. As we discussed in
 Sect.~\ref{sec:correp},  if the sets of replicas $\{{f_a^p}^{(r)}\}$, 
$\{{f_b^p}^{(r)}\}$ that are being combined in Eq.~(\ref{eq:Fdefcor}) are
 randomly selected from the two sets, then the correlation vanishes
 regardless of its true value, see Eq.~(\ref{eq:corravz}). If one
 selects replicas ${f_a^p}^{(r)}$, 
${f_a^p}^{(r)}$  that are fitted to the same underlying data replica,
 then the S-correlation does not vanish, but it is generally underestimated
 because it only includes its data-driven component, as explicitly shown in
 Sect.~\ref{sec:dcorr}.

 The problem is especially severe when combining $n$ different
 sets, because in this case underestimating the correlation between
 each pair of sets will lead
 to an increasingly large underestimation of the uncertainty on the
 combination as the number of sets increases.  This is because in this
 case one is effectively assuming that the differences between the
 different determinations are due to each of them being a partly
 independent measurement, and as such doing more and more determinations
 reduces the uncertainty indefinitely.

\begin{figure}[t]
  \begin{center}
    \includegraphics[width=0.49\textwidth]{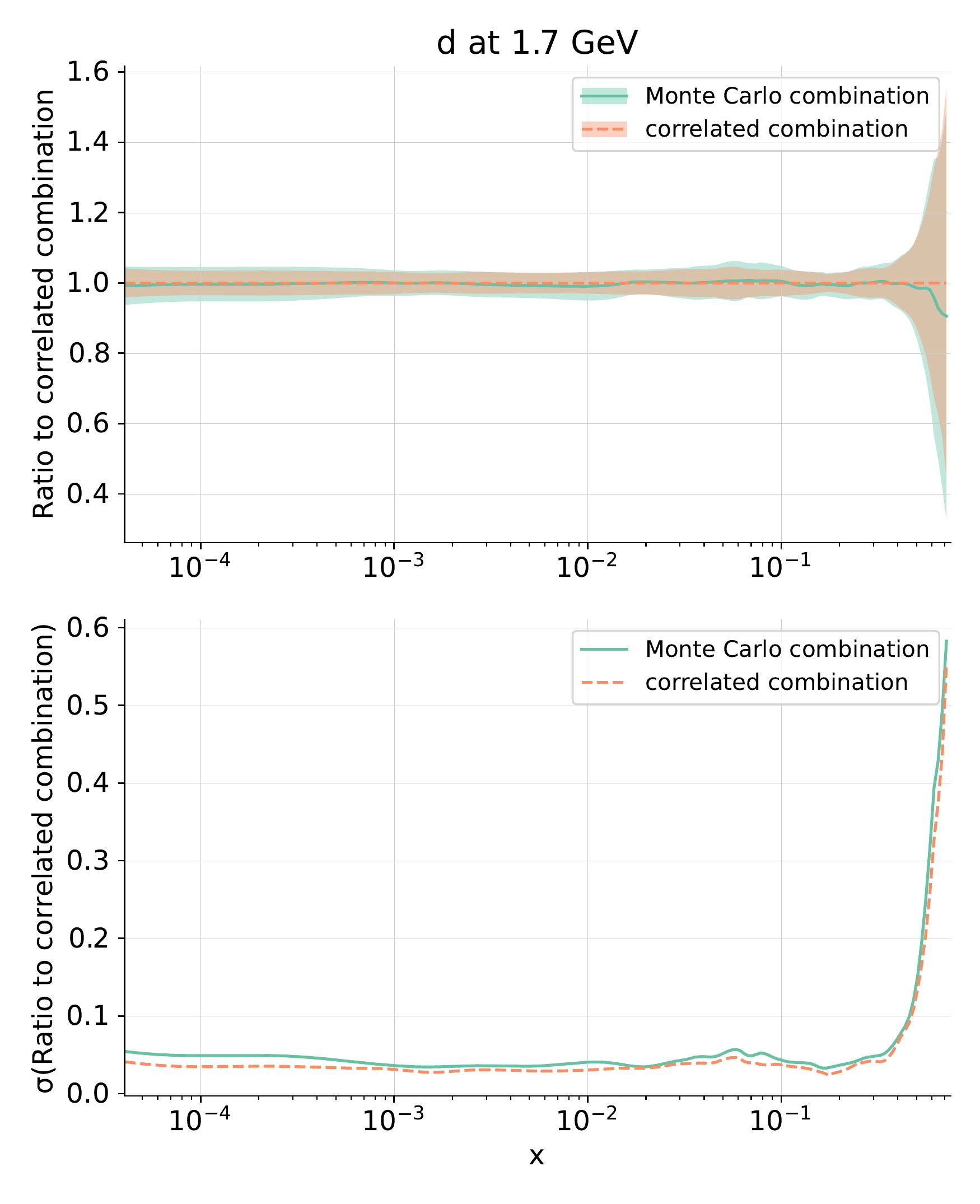}%
    \includegraphics[width=0.49\textwidth]{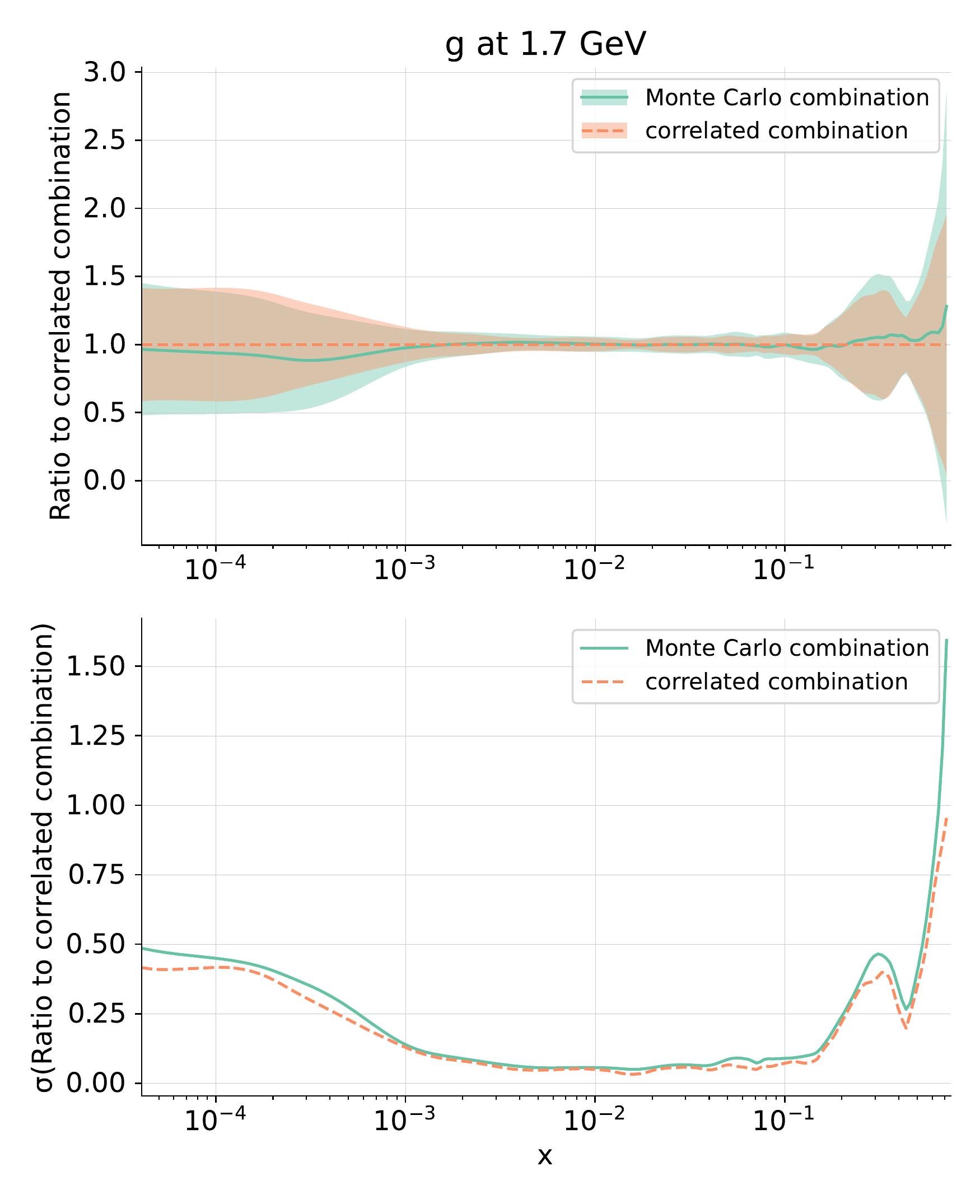}
    \caption{\small Check that the Monte Carlo combination Eq.~(\ref{eq:Fdefcor})  and the
      correlated combination Eq.~(\ref{eq:EFcowan}) of PDF sets yield the same
      answer. Ten sets 
      sets of 43 NNPDF4.0 PDF replicas are combined : 1) as the correlated
      weighted average Eq.~(\ref{eq:EFcowan}) of the ten result from
      the ten sets determined 
      using the data-driven component
      Eq.~(\ref{eq:xcov}) of the S-correlation (correlated
      combination); or 2)
      as the
      average set of 43 replicas obtained using Eq.~(\ref{eq:ECovFncor})  from the ten replicas
      determined from each data replica (Monte Carlo combination).
      Results are shown
      for the down (left) and gluon (right) PDF. We show the PDFs normalized
      to the correlated weighted PDF (top), and the relative 1$\sigma$
      uncertainty (bottom).
      \label{fig:comparecorr}}
  \end{center}
\end{figure}

\begin{figure}[t]
\begin{center}
  \includegraphics[width=0.49\textwidth]{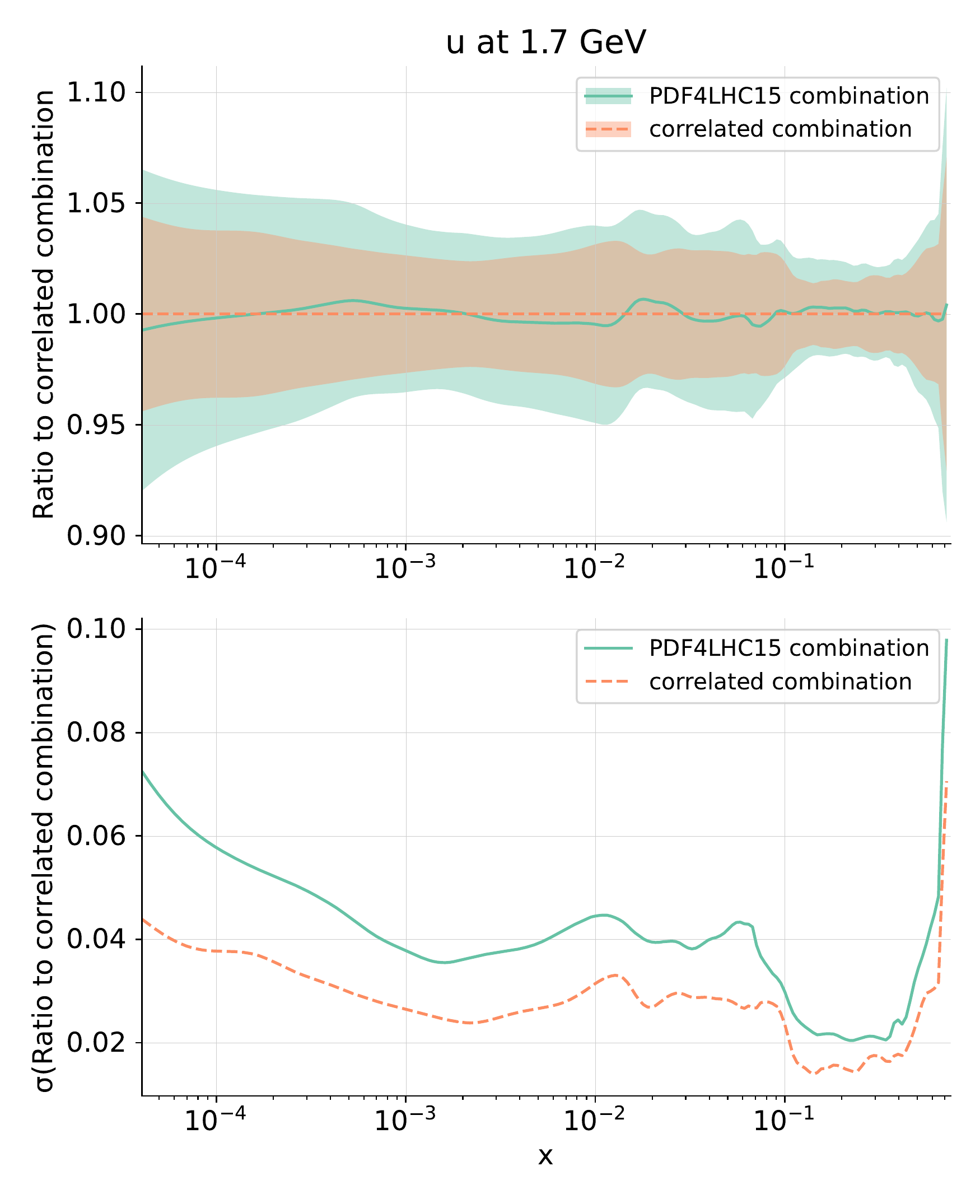}%
  \includegraphics[width=0.49\textwidth]{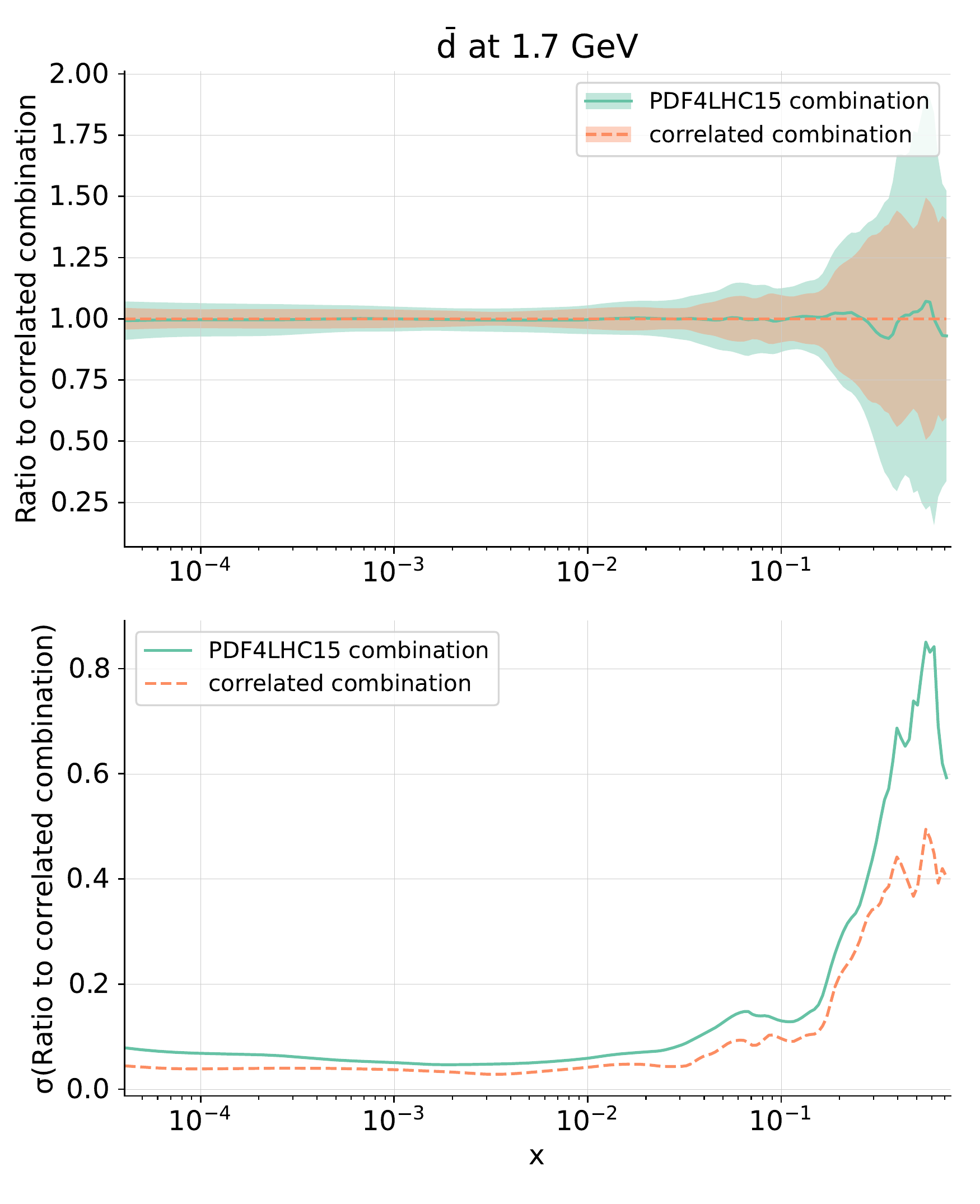}\\
  \includegraphics[width=0.49\textwidth]{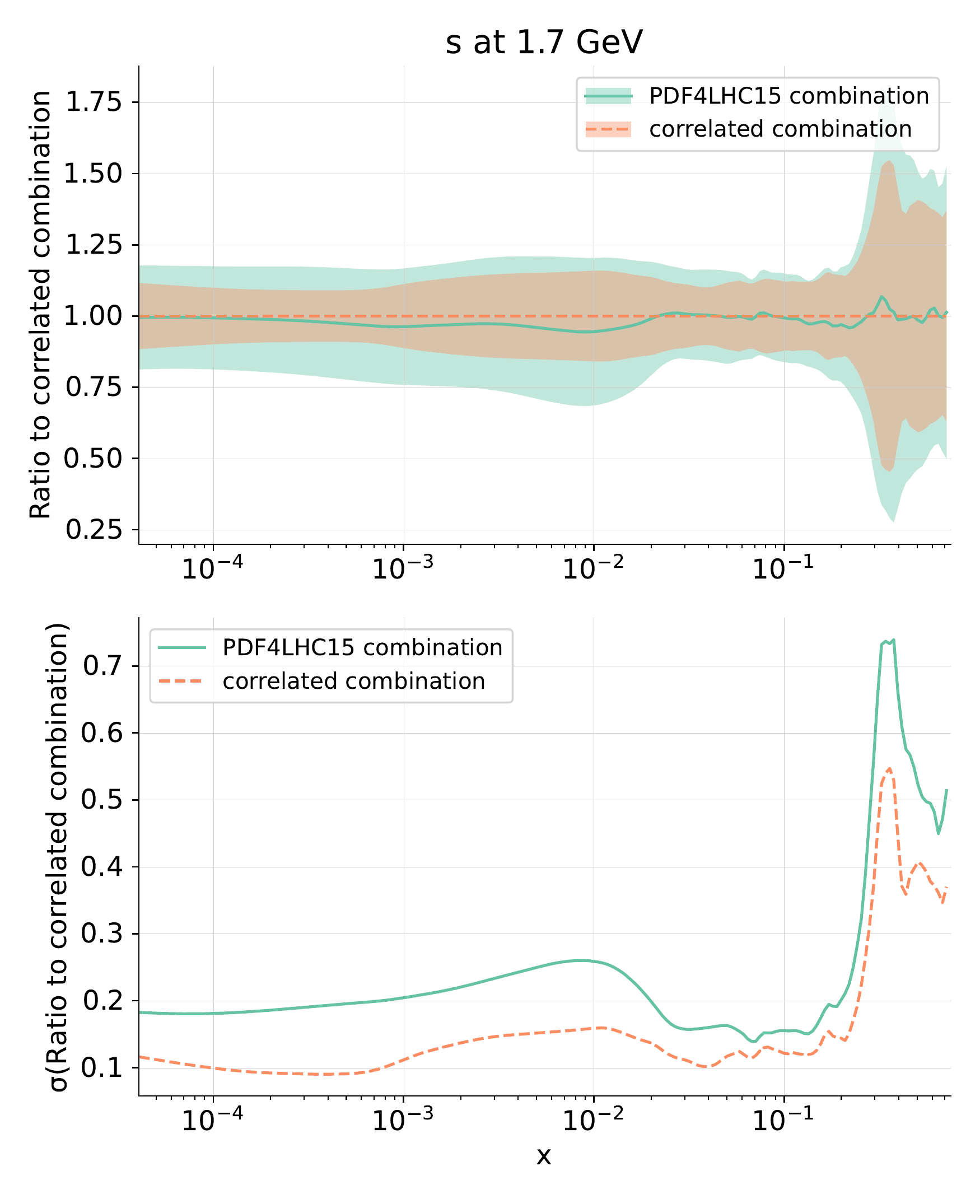}%
  \includegraphics[width=0.49\textwidth]{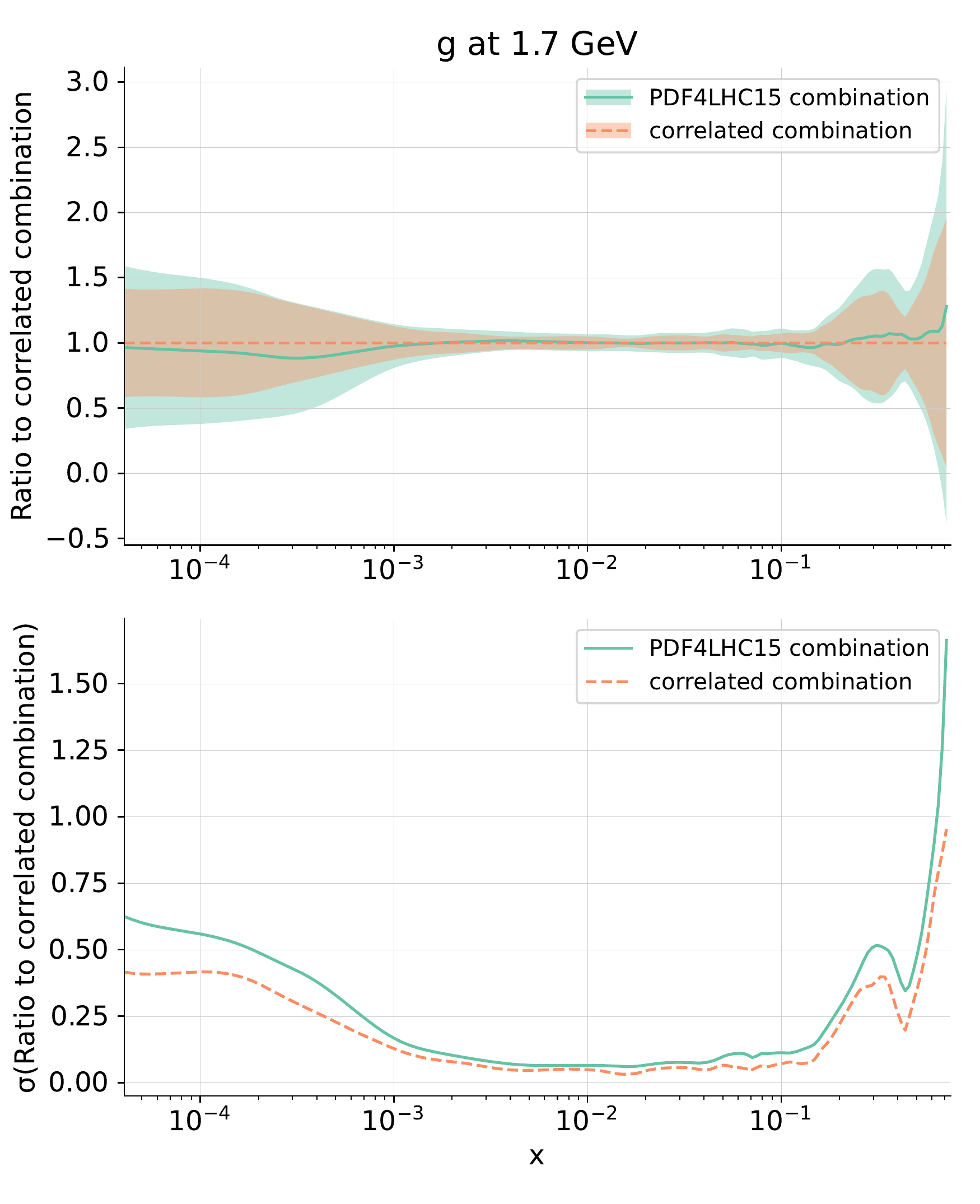}
  \caption{\small Comparison of the uncertainty on the correlated PDF
    combination  and the
    PDF4LHC15 combination  for 10
    sets of 43 PDF NNPDF4.0 replicas. The correlated combination is 
    obtained as the correlated weighted
    average of the ten results Eq.~(\ref{eq:EFcowan}) (same as shown in
    Fig.~\ref{fig:comparecorr}); the PDF4LHC15 combination is found by simply
    combining all replicas in a single 430-replica set.
    \label{fig:coruncor}}
\end{center}
\end{figure}

\clearpage

In order to expose the problem, we have considered an implementation of the combination
Eq.~(\ref{eq:EFcowan}).
We have constructed ten
sets of $N_{\rm rep}$ PDF replicas, all determined from the same $N_{\rm rep}$ underlying
data replicas. In practice we take $N_{\rm rep}=43$ because this is
the largest number we got after applying the procedure discussed in
Appendix~\ref{app:correp}. 
We have then computed the ten by ten S-correlation matrix
Eq.~(\ref{eq:xcov}) for each PDF and each $x$ value,
and we have combined the ten sets  using Eq.~(\ref{eq:EFcowan}).
We have explicitly checked that this is
equivalent to instead using Eq.~(\ref{eq:ECovFncor}) to combine the ten
sets in a single set with 43 replicas. This is demonstrated in
Fig.~\ref{fig:comparecorr} where  two
representative PDFs determined using either method are compared and
seen to agree. This shows that the correlated Monte
Carlo combination  Eq.~(\ref{eq:ECovFncor}) is equivalent to the
combination using the correlation matrix Eq.~(\ref{eq:EFcowan}).

We have then compared this correlated combination to the PDF4LHC15
combination. The latter of course simply consists of
putting together all replicas in a single 430 replica PDF set. Results
are shown in Fig.~\ref{fig:coruncor}. It is evident that while the PDF4LHC15
combination 
gives by construction the correct answer (since in this case
the PDF is simply being combined to itself), 
the correlated combination leads to a rather smaller uncertainty.
Clearly this is absurd. The reduction in uncertainty is the
consequence of the fact that the S-correlation computed using
Eq.~(\ref{eq:xcov}) only includes the data-induced
component. This underestimates the true correlation, because as we
have seen in Sect.~\ref{sec:dcorr} (see in particular
Fig.~\ref{fig:selfcorr}) it leads to a S-correlation which is
rather lower than one, while in actual fact all these PDFs are fully
correlated. The uncertainty reduction is amplified by having combined
ten different sets.

We thus see that combining PDFs determined from the same underlying
data as if they were correlated measurements leads to an incorrect
answer because it neglects the fact that a sizable component of the PDF
correlation is not data-driven. Indeed, if one did determine PDF
uncertainties in this way, one would reach the paradoxical conclusion
that PDF uncertainties can be made smaller at will by simply repeating
many times the PDF determination with the same underlying data
replicas.

\begin{figure}[t]
\begin{center}
  \includegraphics[width=0.5\textwidth]{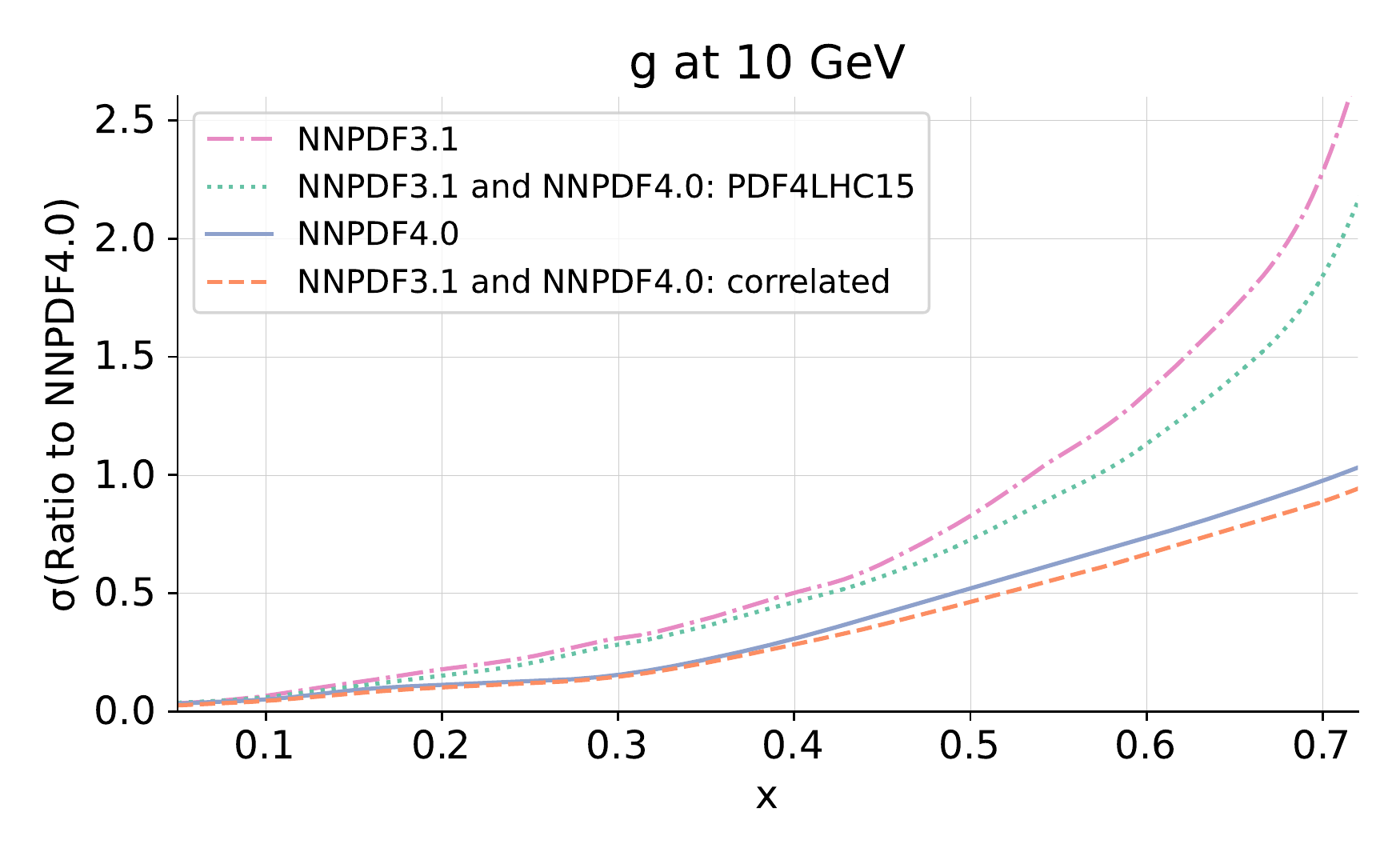}%
  \caption{\small Comparison of the relative uncertainty on the large
    $x$ gluon
    PDF determined using  (from top to bottom) NNPDF3.1 methodology
    (purple, dot-dashed),
    NNPDF3.1 and NNPDF4.0 uncorrelated (PDF4LHC15) combination (green,
    dotted), NNPDF4.0 methodology (blue, solid),
    correlated combination (orange, dashes).
    \label{fig:coruncorg}}
\end{center}
\end{figure}

Because of
the difficulty in accurately estimating the non-data-driven component
of the self-correlation, which is generally significant, this will be
the generic scenario. As an especially striking example of
this situation, in Fig.~\ref{fig:coruncorg} we compare the relative
uncertainty on the gluon PDF that we find if the gluon is determined
using the NNPDF4.0 methodology, the NNPDF3.1 methodology, or the uncorrelated
(PDF4LHC15) or correlated combination. As already discussed (see Fig.~\ref{fig:glucomp}), the
uncertainty found using the NNPDF3.1 methodology is rather larger than that
found using the NNPDF4.0 methodology. Because, as also discussed, central
values are very close, the uncertainty of the PDF4LHC15 combination,
Eq.~(\ref{eq:CovF}) is essentially the average of the uncertainties
with the two methodologies. However, the uncertainty on the correlated
combination is actually smaller than either of the uncertainties with
the two methodologies that are being combined. One would thus reach the
paradoxical conclusion that combining PDFs obtained with the  more
precise NNPDF4.0 methodology with the previous less precise NNPDF3.1
would actually lead to a reduction in uncertainty.

We conclude that a correlated
combination inevitably leads to uncertainty underestimation and it
cannot be considered as an alternative to the  PDF4LHC15
combination, even though the latter might lead to uncertainties that
are a little conservative.

\section{Conclusions}
\label{sec:sum}

Treating PDF sets as independent measurements of the same underlying quantity
provides a new window into the properties of PDFs, specifically the efficiency
of the methodology used to determine the PDFs, and it has allowed us to address
in a quantitative way the possibility of performing correlated
combinations of PDF sets.

We have studied correlations between pairs of PDF determinations,
which we call PDF cross-correlations. By computing these
cross-correlations for PDF replicas fitted to 
the same underlying data replicas,  we have shown
that the correlation of a PDF set to itself does not come out to be
100\% if only the part of the correlation which is due to the
underlying data is included. This is due to the fact that a component
of the PDF uncertainty, and thus also of the PDF correlation, is not
due to the underlying data, but rather of functional origin, namely,
related to the fact that the data do not determine uniquely the PDFs. This realization is what allows
us to use cross-correlations as a tool in the assessment of the efficiency of
the methodology. Indeed, a higher data-driven
cross-correlation is an indication of a more efficient methodology,
i.e. a methodology that correlates more strongly the PDF to the
underlying data.

We have furthermore addressed the feasibility of combining PDF sets as
correlated measurements. Specifically, we showed that combining PDFs as
correlated measurements will lead the uncertainty of the combination to be
underestimated whenever the PDF cross-correlation is
underestimated. This will typically happen
unless it is possible to determine its functional, non-data-driven component.

In this paper we focused on PDFs produced using the NNPDF framework,
in which neural networks are used as   basic underlying interpolating
functions. The neural network parametrization is redundant, and a
best-fit is obtained by not aiming for the minimum of a figure of
merit, which would correspond to overlearning, but rather using a
variety of techniques such as cross-validation to determine the optimal best fit.
In this approach, the presence of a component of the PDF
covariance matrix which is not due to the underlying data follows from
the fact that this best fit is not unique: an
ensemble of best-fit PDFs with non-vanishing covariance is found even
when repeatedly fitting the same underlying data.

However, many other modern PDF sets (e.g. CT18~\cite{Hou:2019efy} and
MSHT20~\cite{Bailey:2020ooq}) rely on parametrizing PDFs with a fixed
functional form, for which a unique best fit to any given set of data
exists, and is given by the minimum of a figure of merit. Note that
this remains true even when used in
conjunction with a Monte Carlo approach, in which the fixed functional form
is fitted to an ensemble of data replicas~\cite{Watt:2012tq}. In this case,
the functional uncertainty is the result of variations of the 
functional form, possibly combined with a  choice of ``tolerance'',
whereby an inflation of the $\chi^2$ variation corresponding to
one-sigma is introduced (either in the data replica generation in a
Monte Carlo approach, or in the fitting in a Hessian approach).

A direct computation of the functional contribution to the covariance
matrix appears to be very difficult when using the NNPDF approach, and
even more challenging when comparing the NNPDF approach to approaches
based on an underlying fixed functional form and tolerance. This is
due to the difficulty in estimating and even more correlating uncertainties which are not
related to the underlying data. In this sense, the results presented
  here are a first step towards a deeper
  understanding of origin and meaning of PDF uncertainties.

  A general lesson to be learned from this exercise appears to be that
  if the aim is to improve the accuracy of PDF determination, it is
  more promising to develop better fitting and diagnostic tools to 
 obtain a more reliable result, than to combine pre-existing
  determinations in the hope of reducing uncertainties
  in the combination.

\section*{Acknowledgments}
We thank all members of the NNPDF collaboration for numerous
discussions, and Alessandro Candido, Juan Rojo and Christopher Schwan
for a careful critical reading of the manuscript and several comments.
SF thanks Simone Amoroso, Daniel Froidevaux and Bogdan Malaescu
for  discussions on PDF
combination.
RDB is supported by the U.K. Science and Technology 
Facilities Council (STFC) grant ST/P000630/1.
SF and RS are supported by the European Research Council under
the European Union's Horizon 2020 research and innovation Programme
(grant agreement n.740006).
\clearpage

\appendix

\section{Computing cross-correlations in the NNPDF framework}
\label{app:correp}

We provide here some details on the computation of the
cross-covariance Eq.~(\ref{eq:xcov}) using NNPDF
methodology.

In the NNPDF methodology the
data replicas are  generated based on a Monte Carlo method with random
initialization.  Furthermore, input data
are split into a training subset used by the optimization algorithm and a
validation subset used to validate the
optimization~\cite{Ball:2014uwa}. This split is performed randomly for
each PDF replica.
In order to compute the data-induced component of the
cross-correlation therefore we have made sure that the two PDF sets
that are being compared are fitted to the same data replicas, with the
same training-validation split.

Furthermore, not all fits end up in the final PDF set, but only those that pass
post-fit criteria specified in Ref.~\cite{Ball:2014uwa}. Because these
criteria are applied a posteriori, it might happen that, for a given
underlying data replica,  the criteria 
are only passed by one of the the two PDF replicas that are being
compared. For the computation of the cross-correlation, we only
include in the final  set PDF replicas for which both sets have passed
the criteria.

The S-covariance is then computed using
Eq.~(\ref{eq:xcov}), or its obvious generalization in
the case of the cross-covariance Eq.~(\ref{eq:selfcovmatpdf}).

In order to estimate the
uncertainty on final results due to the finite size of the replica
sample we have used  a
bootstrapping method~\cite{Efron:1979bxm,Efron:1986hys}. Specifically, we apply a Monte Carlo
algorithm to perform a resampling ``with replacement'' of the PDF replicas.
This is of course done synchronously for both PDF sets between which the
correlation is calculated. We then calculate the PDF correlation using
these resampled PDF sets. This routine is repeated many times to obtain a
precise estimate of the standard error of the PDF correlation. The magnitude
of the uncertainty decreases with the inverse square root of the number of PDF
replicas used to determine the PDF correlation.
We have performed this procedure with 200 resampled sets.
The value was  chosen comparing the 2$\sigma$ standard deviation and the 95\% confidence
 interval, and checking that for any flavor and any value of $x$ they agree.

Finally, we have by default computed the cross-correlation at the
scale $Q_0=1.7$~GeV, and we have checked that by repeating the
computation with different choices of  $Q_0$
up to 100 GeV results are unchanged.

\bibliographystyle{UTPstyle}
\bibliography{corr}

\end{document}